\documentstyle[aps,epsf,epsfig]{revtex}
\def\beq{\begin{equation}}
\def\eeq{\end{equation}}
\def\beqa{\begin{eqnarray}}
\def\eeqa{\end{eqnarray}}
\def\l{\left} 
\def\r{\right}
\def\bdi{\begin{displaymath}}
\def\edi{\end{displaymath}}

\begin{document}
%\tightenlines

\title{Physics of Thick Polymers}

\author{Davide Marenduzzo$^1$, Alessandro Flammini$^2$, Antonio 
Trovato$^2$, Jayanth R. Banavar$^3$ and 
Amos Maritan$^2$ }

\address{$^1$ Department of Physics, Theoretical Physics,
University of Oxford, 1 Keble Road, Oxford OX1 3NP, England}

\address{$^2$ INFM and Dipartimento di Fisica ``G. Galilei'', Universit\a` 
di Padova, Via Marzolo 8, 35131 Padova, Italy}

\address{$^3$ Department of Physics,  104 Davey Laboratory, The 
Pennsylvania  State  University,  University Park,  Pennsylvania
16802, USA}

%\address{$^4$ The Abdus Salam Centre for Theoretical Physics, Strada 
%Costiera 11, 34014 Trieste, Italy}

\vskip 0.5cm

\maketitle

\noindent {\bf Correspondence:} \\
Davide Marenduzzo, Department of Physics, Oxford University,\\
\small 1 Keble Road, Oxford OX1 3NP, United Kingdom,\\
FAX: +44 1865 273947 --- e-mail: d.Marenduzzo1@physics.ox.ac.uk\\

\noindent {\bf Authors' e-mail addresses:} \\
Davide Marenduzzo: davide@thphys.ox.ac.uk\\
Alessandro Flammini: flammini@shannon.sissa.it\\
Antonio Trovato: trovato@pd.infn.it\\
Jayanth R. Banavar: jayanth@phys.psu.edu\\
Amos Maritan: maritan@shannon.sissa.it

\newpage

\begin{abstract}

We present the results of analytic calculations and numerical
simulations of the behaviour of a new class of chain molecules which
we call thick polymers.  The concept of the thickness of such a
polymer, viewed as a tube, is encapsulated by a special three body
interaction and impacts on the behaviour both locally and non-locally.
When thick polymers undergo compaction due to an attractive
self-interaction, we find a new type of phase transition between a
compact phase and a swollen phase at zero temperature on increasing
the thickness.  In the vicinity of this transition, short tubes form
space filling helices and sheets as observed in protein native state
structures.  Upon increasing the chain length, or the number of
chains, we numerically find a crossover from secondary structure
motifs to a quite distinct class of structures akin to the
semi-crystalline phase of polymers or amyloid fibers in polypeptides.

\end{abstract}

%\maketitle

%\pacs{PACS numbers: XXXXX}

\section{Introduction}

Theoretical studies of polymer chains have a long history. Our focus,
in this paper, is the study of a class of polymers characterized by a
non-zero thickness and that can be viewed as tubes similar to garden
hoses or flexible strands of spaghetti. Examples of such polymers
abound and include vital biomolecules such as proteins and DNA.

Classic polymers are different from proteins on several
counts. First, proteins are often very short chains made up of around a 100
or so aminoacids. Second, the aminoacid specificity plays a key role
in the choice of its native state or ground state structure. One can
imagine that the behavior of short chain molecules ought to be
non-universal with the details mattering a great deal. To quote from
Flory\cite{flory_book}, 
{\it ``Synthetic analogs of globular proteins are unknown. The
capability of adopting a dense globular configuration stabilized by
self-interactions and of transforming reversibly to the random coil
are peculiar to the chain molecules of globular proteins alone''.}

In spite of the difficulties mentioned above, a study of the
experimental data on proteins reveal an astonishing simplicity. All
small globular proteins adopt, as their native conformations,
structures made up of simple motifs such as helices, hairpins and
sheets connected together by tight turns\cite{Pauling1,Pauling2}. 
Furthermore, the total number of distinct native state folds
total just a few thousand in
all instead of the vastly larger number that one would expect for a
conventional chain molecule of this length\cite{chothia}. The structures are
flexible and allow for a dizzying array of tasks that enzymes
perform. One might ask where these common attributes of proteins
originate from. Recent work has shown that the concept of a thick
polymer might be useful for bridging the gap between polymer physics
and the biomolecular phase\cite{stasiak,Maritan,rmp,Banavar,tubes,pnas,pre}.

Here we study a thick polymer through numerical simulations and
approximate mean field theory to understand its phase behavior. Our
work is somewhat limited in scope because we do not consider certain
features such as twist rigidity, which are essential for
understanding DNA elasticity\cite{mezard,nelson}. 
It has been suggested that the
effects of twist rigidity are not important for
proteins\cite{Dill}. Furthermore, as we shall demonstrate, the notion
of thickness is sufficient to interpolate between the conventional
compact polymer phase and the phase employed by nature to house
protein native state structures. While non-universal behavior is
expected for short chains, short chains with the right thickness
exhibit very novel finite size effects with some quite robust features
independent of the details. Our results are in excellent accord with
experiments carried out on proteins over several decades and provide a
framework for understanding the common character of proteins.

The simplest paradigm of a polymer consists of spheres tethered
together to form a chain. In the continuum limit, such a model is
described by the classic Edwards model \cite{DOI} which captures
self-avoidance by means of singular delta-function interactions.
However, such a singular interaction does not admit a description of a
chain of non-zero thickness and, indeed, one must renormalize it to
analyze such a continuum model. Furthermore, the Edwards model is
dynamically unable to preserve the knotting number of a closed
chain. Indeed this result arises because the penalty for chain
crossing is finite and thus crossing is not entirely forbidden. The
simple model of tethered spheres is merely a starting point and it has
been modified to take into account local bending energies \cite{flory}
and other features that are known to play a role in experiments.

Here, we take a fresh look at the very basis of the description of a
polymer chain. We suggest that a model of an isotropic sphere as the
basic tethered unit leaves out a crucial feature -- the inherent
anisotropy associated with each unit arising from a special direction
defined by the neighbouring units along the chain. Physically, we
argue that this anisotropy can be captured by considering a chain made
up of tethered coins or discs, which in turn leads to a tube picture of the
chain.  From a theoretical point of view, the model of tethered coins
has the non-trivial advantage of lending itself, in a natural and
singularity-free manner, to a continuum description. The singular
potential is avoided because one can imagine stacking together thinner
and thinner coins closer and closer together while maintaining the
coin radius constant to obtain a tube of non-zero
thickness\cite{Buck}.

One may rationalize the description of a generic polymer such as
a polyethylene chain by a thick tube. Polyethylene chains have a Kuhn
length around $1$ nm and an approximate hard core diameter of a united
atom of methylene around $0.35$ nm\cite{polyethylene}. The picture of a tube
of non-zero thickness is thus a physically motivated description of
such a chain and can serve as an alternative to the model of a
flexible line with a $\delta$-function hard core self-avoidance
interaction. Indeed, the key difference in the two cases is in the
nature of self-avoidance.

One elegant way of describing the self-avoidance of a tube is by means of
a three body potential, which enforces the constraint that a suitable
length, constructed in terms of the positions of any triplet of points
sitting on the axis of the tube, is greater than the tube thickness.
As explained in Section III, this length associated with a triplet of
points is the radius of the circle through them\cite{K1}.  This length
is the local radius of curvature when the three points tend to a
single point on the curve, but non-local triplets play a significant
role as well. The allowed configurations of a simple system of unthetered
hard spheres are those in which one considers all pairs of spheres and
for each pair, one ensures that the sphere centers are no closer than
a sphere diameter. Such a rule ensures that the spheres are
non-overlapping.  The generalization of this to the chain context is
to consider all triplets along the axis of the tube. Self-avoiding
conformations of a flexible tube are obtained by drawing circles
through each of the triplets and ensuring that none of the radii is
less than the tube radius.

\begin{figure}
\centerline{\psfig{figure=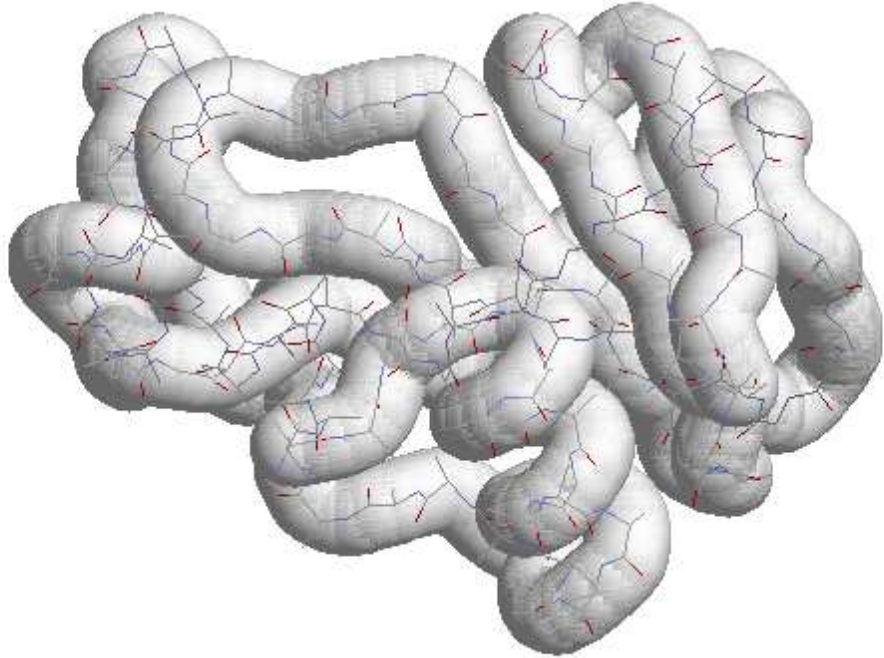,width=6.in}}
\caption{An all atom representation of a segment of protein backbone 
(thin colored lines) together with its corresponding backbone 'tube'.
The backbone $C_{\alpha}$ atoms and the 
side chains of the amino acids impose steric constraints which leads to
an effective tube of non-zero thickness.
The tube is not inflated up to its maximum thickness for convenience
of visualization.}
\label{tubo}
\end{figure}

A tube description is useful for the understanding of proteins -- the
backbone of a protein, when viewed as a tube, has an intrinsic
non-zero thickness in order to accommodate the steric constraints
imposed by the atoms of the side chains (for a sketch, see
Fig. \ref{tubo}). Our model is characterized by two parameters other
than the length of the tube, its thickness $R_0$ and the range of the
attractive potential $R_1$. When these two lengths are comparable, we
find that the ground state configurations of a short tube closely
resemble the secondary motifs of folded proteins.  We will demonstrate
that these configurations change to amyloid-like structures upon
increasing the tube length.  We will present detailed studies of the
phase diagram and will discuss similarities and differences with those
obtained with previously studied polymer models. Through an
extensive statistical analysis of triplet radii in known ground states
of proteins, we found that the average thickness of a protein is
$0.27$ nm\cite{Banavar}. Consequently, the typical values of $R_0$ and of 
$R_1$ are matched in native structures of proteins, where $R_1\simeq 0.55$ nm,
when intra-chain hydrogen bonding is considered.

%The novelty here is that crucial features of biopolymers simply emerge
%due to the interplay between the thickness and the range of attractive
%forces on monomers. This is not to say that we ignore the complex
%chemical origin of such forces, but that, to a large extent, it is the
%range of their cumulative action and how it relates to the size of the
%polymer that counts.  It is also important to stress that all results
%presented here are largely independent of the particular
%implementation of the thickness we choosed, and therefore they are not
%an artificial by-product of the 3-body interaction\cite{K1} by which we model
%the thickness itself.

There are several new features, both at local and non-local levels,
that are characteristic of a tube.  First, there is a constraint on
the local radius of curvature which forbids the radius from being
smaller than the tube thickness. This constraint is different from the
bending rigidity potential energy \cite{flory} which is usually taken
to be of the form $\kappa(1-\cos(\theta))$ where $\kappa$ is a
positive coupling, and $\theta$ the angle between successive links --
as the temperature is lowered the chain becomes straighter. In
contrast, the local radius constraint in a tube is a rigid constraint
which is independent of temperature. In addition, the tube imposes a
non-local constraint which, physically, reflects the fact that
different segments of a tube cannot approach each other closer than a
threshold distance governed by the tube thickness.  This is captured
by imposing that no three-body radius (of any non-local triplets) can
be smaller than the tube thickness.  In order to best avail of the
attractive interactions between different parts of a tube, the
relative orientation of neighbouring segments becomes important
reflecting the inherent anisotropy associated with a tube
\cite{note11,note12}.
 
These new features lead to a phase diagram in which one can distinguish
between three classes of tube thickness.
In the thin tube regime, one obtains two distinct phase transitions on 
lowering the temperature from the swollen phase. First one enters an
isotropic collapsed 
phase which is supplanted by an oriented nematic-like phase after the second 
phase transition. For intermediate thickness, there is only one phase 
transition to an oriented, compact phase while, for larger tube sizes, 
one simply obtains the swollen phase. 
%We compare and contrast our results with those 
%obtained for a semiflexible chain subject to 
%attraction\cite{stiffness,muthukumar,lise}.

Our studies of short tubes in the intermediate thickness regime show
several unusual characteristics. First, the number of ground state
structures is much smaller than the corresponding number for chains of
spheres: the energy landscape is vastly simpler. Second, the resulting
structures are marginally compact (the effects of attractive
self-interactions have just set in) and, because of their proximity to
a phase transition, are sensitive to the right types of
perturbations. Strikingly, we find that the resulting structures are
predominantly space-filling helices (with a specific pitch to radius
ratio) and nearly planar zig-zag hairpin and sheets. These structures
are found to be quantitatively akin to the building blocks of protein
structures.  Our results indicate the presence of a previously
unstudied phase of matter, associated with short tubes of intermediate
thickness, which has many attendant advantages exploited by nature in
housing biomolecular structures.  Our results show that the tube
picture provides a unified framework for understanding the common
character of small globular proteins.  Upon increasing the length of
the polymer, or the number of polymer chains, we observe, in computer
simulations, a crossover to semi-crystalline structures with different
portions of the backbone chain lying parallel to one
another. Significantly, this low temperature anisotropic phase of
tubes provides a simple rationalization for the formation of amyloid
in misfolded proteins (leading to deadly diseases, including
Alzheimer's and the Mad Cow disease\cite{PRUSINER}) and the formation
of semicrystalline polymer.

There are some remarkable differences between the behavior of our
model and that of a stiff chain as embodied by the semiflexible chain
model\cite{stiffness,muthukumar,lise}.  The phase diagrams of a
self-interacting tube and a semiflexible chain both display three
phases: a swollen phase, an isotropic collapsed phase, and an
anisotropic globular phase (semi-crystalline phase). However, the
phase boundaries look rather different. In particular, there is a
zero-temperature first order phase transition in the thick polymer
case which is observed on increasing the thickness, which has no
counterpart in the stiff polymer phase diagram.  Moreover, the
thickness acts as a constraint on viable configurations, while the
bending rigidity acts as an energetic penalty. Technically, this leads
to the swollen phases being somewhat different in that in the
continuum limit the typical tube centerlines are smoother than the
semiflexible chain backbones.  Also short tubes of intermediate
thickness yield helices and almost planar zig-zag conformations, a
feature absent in the semi-flexible chain model.

Our paper is organized as follows.
Section II presents a brief review of the Edwards model.
Section III presents the tube picture of a polymer chain  and its
advantages. Section IV
deals with studies of the ground states of short chains subject to 
self-attraction and underscores the qualitative differences between chains of
spheres (with or without bending energies) and short tubes. We review work 
which shows that the latter describes a novel phase used by nature to house 
biomolecular structures. Section V presents an attack on deducing the phase 
diagram of tubes using several complementary techniques. Some of the 
technical details are relegated to an appendix. We conclude with a brief 
summary in Section VI. 

\section{Edwards model of a polymer chain} 

Field theory models have proved to be useful for carrying out analytic
calculations of the scaling behavior of polymers.  The standard model is
that due to Edwards -- the polymer is described 
by a continuous curve, $\vec{r}(s)$, with an effective energy 
given by
{\setlength\arraycolsep{2pt}
\begin{eqnarray} 
H(\{\vec{r}\}) & = & \frac{1}{2} \int_0^L \dot{\vec{r}}(s)^2 ds 
 +\frac{v_2}{6} \int_0^L\int_0^L  \delta(\vec{r}(s)-\vec{r}(s'))ds ds' +
{} \nonumber\\
& & {}+ \frac{v_3}{90} \int_0^L\int_0^L\int_0^L 
\delta(\vec{r}(s)-\vec{r}(s')) \delta(\vec{r}(s)-\vec{r}(s''))dsds'ds''
+ {} \cdots \label{ham} 
\end{eqnarray}}

The first term in the above equation takes into account the chain
entropy and arises from the central limit theorem.   It is readily 
derived for a non-interacting chain of beads (the location of the
$i$-th bead is denoted by $\vec{r}_i$) tethered
together by a potential $u(r_{j,j+1})$, with 
$r_{m,n}=\|\vec{r}_n -\vec{r}_m\|$, acting between adjacent beads
and keeping them at a typical mean square distance $b^2$. 
Using the central limit theorem,  one finds
that the probability of finding the first bead at $\vec{r}_1$ and
the $k$-th bead at $\vec{r}_k$ is well approximated by 

\begin{equation}
P(\vec{r}_1,\vec{r}_k) \propto
\exp\{-\frac{d}{2b^2k}(\vec{r}_1-\vec{r}_k)^2\} 
\label{cent} 
\end{equation}
for large $k$.
Consider now  a `coarse-grained' chain made up of 
$n$ pieces  of such $k$-step chains.   Let the $i$-th piece start at the position
$\vec{r}_{(i-1)k}$  and end at  $\vec{r}_{ik}$  for each
$i=1,2,...,n$.  It then follows that the probability of obtaining this
configuration of the chain  is given by

\begin{equation}
P(\vec{r}_0,\vec{r}_1,...,\vec{r}_n) \propto
\exp\{-\frac{d}{2b^2k}\sum_{i=1}^n(\vec{r}_{(i-1)k}-\vec{r}_{ik})^2\}  
\label{prob}
\end{equation}

In the continuum limit, one defines $\epsilon=kb^2/d$ and  the total
`length' of the `coarse grained' chain, $L=n\epsilon$, is kept fixed
while $n\to \infty$ (this implies that $b \rightarrow 0$).   In this limit,
the previous equation leads to the Hamiltonian 
(Eq. (\ref{ham})), in the non-interacting case ($v_i=0, i=2,3,...$), because

\begin{equation}
\lim_{\epsilon \rightarrow 0}
\sum_{i=1}^n(\vec{r}(t_{i-1})-\vec{r}(t_{i}))^2/\epsilon  =
\int_0^L \dot{\vec{r}}(s)^2 ds 
\end{equation}
where $\vec{r}_{ik}=\vec{r}(t_i)$.\\

In the self interacting case, one again starts from a string of beads as
before, where the total energy is given by,
\begin{equation}
H_{chain}=\sum_{j}u(r_{j,j+1}) + \sum_{m<n}V(r_{m,n})
\label{hchain}
\end{equation}
where $V(r)$ is a two-body interaction between pairs of beads at a
distance $r$. The shape of $V(r)$  is typically similar to the classic
6-12 Lennard-Jones potential with attraction at intermediate length scales
and repulsion when two beads get too close to each other.
A link between Eq.  (\ref{hchain}) and  the continuum model Eq. (\ref{ham}) 
can be established through the  virial expansion. \\

The $v_m$'s in eq.(\ref{ham}) represent effective
$m$-body interactions for the continuum chain, $\vec{r}(s)$, and they
depend on the temperature, $T$.   In order to get a swollen phase at
high temperatures and a compact phase at low temperatures, 
at high (low) $T$, $v_2 > 0$ ($v_2 < 0$)
whereas $v_3 > 0$. The three-body term is strictly necessary only in
the low temperature region in order to stabilize the system. 
The model
defined by eq.(\ref{ham}) is widely used in the literature and it has
been shown to have a precise meaning in perturbation theory using
renormalization group techniques (see \cite{CLOISEAUX,YAMAKAWA,GENNES} 
and the original
references therein) \cite{PHI4}.  However, it has been suggested by 
Barrett and 
Domb \cite{domb} that {\em the renormalization group has not provided
a satisfactory description of a polymer in the poor-solvent (weak 
coupling) regime, perturbation methods are hopelessly incapable of 
describing the good solvent (strong-coupling) regime, there is no 
entirely satisfactory description of the crossover from poor 
solvent to good solvent conditions, and the limits in which excluded-volume
chains exhibit universality are not fully understood.} \\

One may understand heuristically why the continuum
version of eq.(\ref{hchain}) can only give rise to interaction terms
which are singular (the $\delta$-function potentials in
eq.(\ref{ham})). Let us make the physical assumption that 
the potential $V(r)$ in eq.(\ref{hchain}) has a
repulsive part which becomes larger as $r$ decreases. In order to take the
continuum limit,  the bead density along the chain must increase, because the
tethering $u$ potential constrains successive beads to lie closer to
each other as 
$b \rightarrow 0$.   As $b$ decreases,  the number of bead-pairs within
the repulsive region of the $V$ potential
increases and, in the continuum limit, the chain would have 
an infinite energy and an infinite rigidity. 
As $b \rightarrow 0$, this unphysical consequence can be avoided
by simultaneously shrinking the size of the repulsive
region of the interaction potential $V$ in eq.(\ref{hchain}) 
leading to singular $\delta$-function potentials in the 
continuum limit as in eq.(\ref{ham}).\\

Furthermore, there does not exist, to our knowledge, a continuum
formulation of closed polymers which allows one to carry out studies
within a given knot class, i.e. with a fixed number of knots. 
Indeed, a drawback of any formulation using 
two-body potentials, such as eq. (\ref{ham})), is that the
self-intersection of the chain is allowed albeit with some energetic
penalty and thus one can change, at will, the topology of the
polymer. This is not what happens in realistic situations.
Furthermore, there are interesting physical problems pertaining to the
estimation of entropic exponents and weights of polymer configurations
within a given topology\cite{Grosberg,stasiak2}. While such
calculations can be carried out numerically, at present, there is no
simple analytic formulation of this problem. \\

A simple discrete model of spheres tethered together in a chain would
lead, in the continuum limit, to the Edwards model.  As the distance
between successive spheres shrinks, so must the radius of the spheres
and in the limit one obtains a string of infinitesimal thickness.
So how would one describe mathematically a string of non-zero thickness
analogous to a tube or a garden hose or a spaghetto?  Is there a way to avoid
a singular interaction potential?  More important, 
is the physics of thick strings
qualitatively different from that predicted by the Edwards model?
Our focus, in this paper, is on answering these questions.

%We will briefly
%review our earlier work which presented a simple prescription of a many
%body potential for describing a tube of non-zero thickness which deftly
%avoids the singularity in the interaction potential.  We will study
%the phase diagram of a tube of non-zero thickness subject to self-avoidance
%constraints and attractive self-interaction.  The physics is enriched 
%with the presence of two length scales, the range of the attractive
%interaction and the tube thickness.  Using approximate analytic calculations
%and numerical simulations, we deduce a novel phase diagram for such a system.
%Our results enable us to describe, within a unified framework,
%well-known polymer phases and a
%novel phase of matter used by Nature to house biomolecular structures.\\

\section{Tube picture of a polymer chain}

In this section, we provide a brief description of how one might describe
a tube of non-zero thickness.
Let us consider the general case when the interacting particles (monomers)
are restricted to lie in D-dimensional manifolds such as a string (D =
1) or a surface (D = 2) embedded in d-dimensional space.  Examples of such
systems include strings and random surfaces and are widely studied in many
branches of science\cite{NELSON,WIESE}.
The interaction (tethering) potential leading to the
system being restricted in the form of a manifold or bending rigidity
terms are not our concern here -- there are many satisfactory ways to
construct such a potential for strings and surfaces.  Our focus is on the
self-interaction not captured by the tethering potential.  This
self-interaction between the monomers are meant to be effective
interactions resulting from the elimination of the finer degrees of
freedom. We will argue that the basic interacting unit must have at least
D+2 monomers in order to define a meaningful characteristic length
associated with this interaction. Such a basic many-body interacting unit
is necessary for a continuum approach to these classes of problems.  Note
that for unconstrained particles, D = 0 and pair-wise interactions
suffice. \\

Consider a D = 1 system of a string  in d = 3.  In order to account for
steric interactions, that prevent the string from intersecting with
itself, one might postulate a pairwise potential which becomes large  when 
two of the constituent particles are near each other.  However, given two
nearby particles, it is impossible to distinguish whether they are from
distinct parts of the string or whether they are close by simply because
of the string connectedness.  As one approaches the continuum limit by
increasing the density of particles along the string and shrinking the
distance between them, the energy contribution from the latter category of
pairs, which are close to each other essentially because of the tethering
potential,  vastly exceeds the contribution from particles that are
genuinely responsible for a self-intersection. 
There is no inherent small-distance cutoff in the
theory and regularization procedures are needed that re-introduce such a
length scale in order to avoid infinities. \\ 

The requirements for a well-founded theory are that one ought to be able
to take a continuum limit on increasing the density of particles, that
self-interactions be properly taken into account and that there be a
characteristic microscopic length other than the spacing between
neighboring particles along the string.  As explained above, a pairwise
potential is not equal to the task.  Let us consider a three-body
potential characterizing the interaction between three particles, which
lie on the corners of a triangle.  Let the sides of the triangle have
magnitudes $r_1$, $r_2$ and $r_3$.  In order to specify a triangle 
uniquely, one
needs three attributes.  The potential of interaction can therefore depend
on three independent length scales, which are invariant under translation,
rotation and permutation of the three particles.  One may choose these length
scales to be the perimeter, P, of the triangle, the ratio of the area, A,
of the triangle to its perimeter, P, and finally $r_1  r_2 r_3 / A$.  The first
two lengths do not cure the problems alluded to before -- they both vanish
when the particles approach each other either
from the same region or different regions of the string. 
The third length scale is proportional to R, the radius of a circle
drawn through the three particles and has proved to be valuable for the
study of knots \cite{stasiak}. \\

If one considers the minimum over all triplets of this quantity,
that is just the `thickness' \cite{K1} of the continuum curve we are
considering -- the thick curve may be thought of as a tube or a thick
polymer. The key result is that the description of the effective
self-interactions of a string, which satisfies all the requirements of
a well-founded theory will involve a three body potential, $V_3(R)$,
among all triplets of particles in the string.  In this case, the
continuum limit can be taken safely, self-avoidance is respected and a
cut-off scale naturally arises through the functional form of $V_3$, in
this case a hard-core form. \\

One may illustrate the difference between the approach to the
continuum limit in the chain of spheres model and the tube model in the
following manner. Consider first a self-avoiding chain made up of spheres
tethered together with the distance between successive spheres
equal to the bond length $b$.  Let $R_{hc}$ represent the hard
core radius of the sphere.  
%Physically, $b$ cannot be less than
%$2 R_{hc}$, because the spheres would overlap.  Also the bond
%length $b$ cannot be greater than $2^{3/2} R_{hc}$ to maintain
%self-avoidance and prevent bonds from overlapping.
Physically, $b$ is expected to be of the order of $2 R_{hc}$.
For a chain of length $N$, one may,
from general considerations, write the radius of gyration as
$R_g(N, b, R_{hc}) = R_{hc} F_s(N/R_{hc},b/R_{hc})$.  In order
to take the continuum limit, one must let $b$ approach 0.  But
this automatically constrains $R_{hc}$ to go to zero as well,
while maintaining the ratio of $b/R_{hc}$ at some suitable
non-zero value.  This immediately poses a problem because, in
order for one to obtain the well-known result that $R_g$ scales
as $N^\nu$, there is no other non-zero length scale in the
problem.  The cure for this, of course, is to use renormalization
procedures by introducing an artificial cutoff followed by a
demonstration that the answers do not depend on the cut-off. \\ 

This problem is nicely avoided in a tube description within which
$R_g(N, b, R_0) = R_0 F_t(N/R_0,b/R_0)$, where $R_0$ is the tube
thickness.  The only constraint on $b$ now is that it is less than or
at most comparable to $R_0$. Thus, the continuum limit can be safely
taken by letting $b$ go to 0 and obtaining $R_g = R_0 F_t(N/R_0,0)
\sim R_0^{1-\nu} N^\nu$.  In other words, the nonzero thickness (as
embodied in a suitable three body potential) of a tube provides a
suitable cut-off length scale and avoids the singularity implicit in
the conventional Edwards model and the absolute need for the use of
renormalization techniques \cite{ZinnJustin}, allowing for the first
time analytic studies of polymers with fixed knotting numbers. In the
continuum limit, the many-body potential replaces the pairwise
self-interaction potential and ought not to be thought of as a higher
order correction\cite{note13}. \\

Nevertheless, for discrete models of a thick tube, two-body
interactions do not suffer from the problem described above and can
coexist with three-body terms which model the non-zero thickness of
the curve approximately.  Consider the Hamiltonian for a string of the
form
\begin{equation}
H_{chain}=\sum_{i}u(r_{i,i+1}) + \sum_{i<j}V_2(r_{i,j})
+ \sum_{i<j<k}V_3(r_{i,j,k}),
\label{h3body}
\end{equation}
where, as before $u$ is a generic tethering potential, $V_2$ is a
pairwise potential and $V_3$ is a three-body potential.  In most
of our simulations, we have chosen $u$ to constrain $r_{i,i+1}$
to be constant.  The pair potential is taken to be
\beq\label{two_body_potential}
V_2(r_{i,j})= 
\left \{ 
\begin{array}{cc}  
 \infty & \mbox{ if $r_{i,j} < 2  R_{h.c.}$ } \\
 -1   & \mbox{ if $ 2 R_{h.c.}< r_{i,j} <  R_{1}$ } \\
  0   & \mbox{ if $R_{1} < r_{i,j}  $ }
\end{array}
\right.  \eeq where $R_1$, the range of the attractive interaction is
taken to be $1.6$ units (measured in units of the bond length or the
fixed distance between successive beads, which is chosen to be $1$) in
our simulations and $R_{h.c.}$ is the hard core radius, which we take
to be $0.55$ in our calculations.  These values have been selected in
order to mimic values typical of a protein backbone (see
Introduction).  We verified that small changes in the attraction range
do not change the results shown here appreciably (see Fig. 4). A
Lennard-Jones potential for the two-body interaction energy, with an
equilibrium distance slightly larger than the neighbouring site
distance, is also expected to give qualitatively similar results,
provided the potential is truncated in order to eliminate the
unphysical effects of a long range tail.  \\

\begin{figure}
\centerline{\psfig{figure=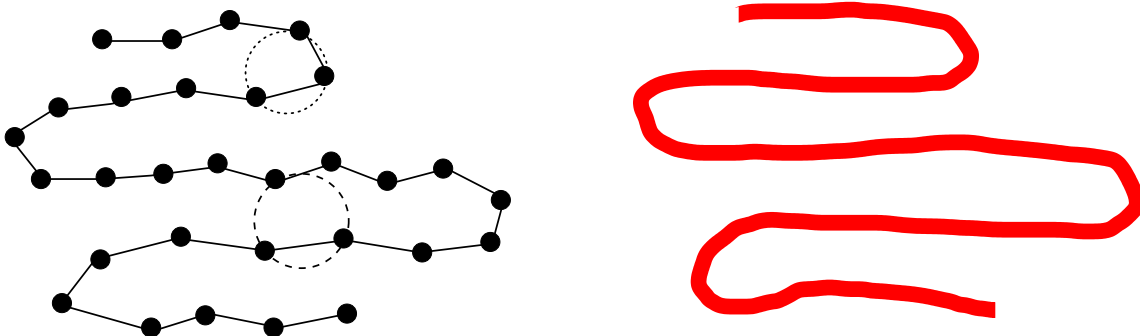,width=6.in}}
\caption{Left: sketch of the backbone of a chain representing a discrete
thick polymer: the black spheres are the position of monomers, the
thickness is imposed by requiring that no triplet radii is bigger
than the required thickness $R_0$: two such radii, one involving 
consecutive points (dotted line) and one non-consecutive points 
(dashed line) are shown. Right: the continuum conunterpart of the
discrete polymer is the red tube drawn here. }
\label{triplets}
\end{figure}

The three body potential \cite{K1,BM}, 
$V_3$,  disallows conformations for which
$R_0>{\rm min}_{i \neq j \neq k}r_{i,j,k}$, where $r_{i,j,k}$ is the 
radius of the circle going through the centers of the beads $i$, $j$ and 
$k$ (Fig. \ref{triplets}).    This three body constraint leads,
in the continuum limit, to a tube of radius $R_0$, whose axis is
given by the string defined by $\vec r_i$ (see Fig. \ref{triplets}). 
This can be simply proved by noting that 
the quantity ${\rm min}_{i \neq j \neq k}r_{i,j,k}$ is the thickness of
the maximally inflated tube with centerline $\{\vec r_{i}\}_{i=1,\ldots,N}$
\cite{K1}. If the maximally inflated tube has a thickness greater than 
$R_0$, then this means that the discrete curve under analysis is a viable 
centerline for a tube of thickness $R_0$.
For the discrete case, the three-body constraint is not as severe as in
the continuum limit --  we will show that 
discreteness can play a vital role in producing planar structures
for short tubes subject to self-attraction.\\

In the remaining
sections, we will demonstrate, using various complementary techniques,
that a rich variety of new phases is obtained for a tube subject to 
self-attraction, which allow a bridging of conventional polymer phases to
a novel phase used by nature for housing biomolecular structures. \\

%Recall that a discrete model of spheres tethered together in a chain would
%lead, in the continuum limit, to the Edwards model.  Physically, the key 
%observation behind the tube model is that it is not
%appropriate to think of isotropic spheres as the building blocks of chains
%because there is a special local direction that one may
%associate with each sphere.
%This local direction is defined by the adjacent spheres along the chain and 
%indicate the local tangent to the chain conformation.  Symmetry considerations play an
%important role in determining the behavior at long length scales
%and it is appropriate to replace spheres by coins in which the head-to-tails 
%direction is distinct from the other two directions.  A chain of stacked
%coins is the appropriate discrete description of a tube of non-zero thickness.
%With a chain of spheres, in which the spheres shrink isotropically,
%as the continuum limit is approached, one obtains an infinitesimally
%thin string in the limiting case.   In contrast, for a chain of coins, 
%one can envision packing thinner coins nearer each other, while yet holding
%their radii fixed.  This would indeed allow one to obtain a tube of non-zero
%thickness in the continuum limit,  avoid the singularity in the.

\section{Ground states of short chains}

Our focus, in this section, is to discuss the ground state structures of
short chains subject to self-attraction.  Strikingly,  marginally
compact tube structures have helical, hairpin and sheet configurations
which are the building blocks of protein native state structures.  We will
consider the structures adopted by a tube and by a chain of spheres in order
to understand the common features of both classes and underscore the qualitatively
new features introduced on incorporating the intrinsic local anisotropy of a chain. 
We will also assess the nature of the structures adopted
by a stiff chain, or equivalently a semiflexible polymer, subject to
compaction. This analysis demonstrates 
that, at least, within the parameter
range we have explored, there is less secondary structure content 
in short semiflexible chains in clear contrast to the
marginally compact phase obtained for a tube.\\

Hard spheres are the simplest basic entities for modelling matter.  In spite of
their simplicity, hard spheres exhibit an entropy-driven phase transition 
between a  fluid phase at low packing fractions
and a crystalline phase at high packing fractions.  The favored crystalline structure
is that of a face-centered-cubic crystal which allows for the most efficient packing.
In two dimensions, hard disks would prefer to pack most efficiently in a triangular
lattice.  Let us now consider a chain molecule made up of tethered hard spheres
(in three dimensions) or tethered disks (in two dimensions).  For simplicity,
let us restrict ourselves to the simplest case in which the tethering constraints
for compact conformations are not frustrating and lead to the same ground state
as in the untethered case.  This greatly simplifies the analysis of the ground state
conformations of compact chains because one can carry out the analysis for
unconstrained, untethered objects first and imagine placing the tethers through 
the resulting ground state structure(s). \\

The systems that we will study (without and with the three body tube
constraint, and with the stiffness term) 
are all subject to the same attractive potential energy
of interaction.  They consist of a chain of spheres subject to a pair-wise
attractive potential -- 
a pair of spheres,
at  $\vec r_i$ and
$\vec r_j$, have an interaction given by eq. \ref{two_body_potential}.
%:
%\beq\label{two_body_potential}
%V(r_{i,j})=
%\left \{
%\begin{array}{cc}
% \infty & \mbox{ if $r_{i,j} < 2  R_{h.c.}$ } \\
% -1   & \mbox{ if $ 2 R_{h.c.}< r_{i,j} <  R_{1}$ } \\
%  0   & \mbox{ if $R_{1} < r_{i,j}  $ }
%\end{array}
%\right.
%\eeq
There is an additional constraint for the tube case
which forbids any of the three-body radii associated with the spheres of the chain
from being smaller than the tube thickness
$R_0$ (potential $V_3$ in eq. (\ref{h3body})). 
Finally, for a 
chain of spheres with a non-zero stiffness \cite{stiffness,muthukumar}, 
we  complement the  
two body potential in Eq. \ref{two_body_potential} with a 
bending-rigidity term of the form:
\begin{equation}
H_b = - \kappa\sum_i \vec r_{i,i+1}\cdot\vec r_{i+1,i+2},
\label{bending_rigidity_term}
\end{equation}
with $\kappa$ a constant, the chain stiffness, and with
$\vec r_{i,i+1}=\vec r_{i+1}
-\vec r_i$. We will point out similarities and differences between these
three cases.

\subsection{Geometry of chains of spheres}

We begin with an analysis of the chain {\em without} the three-body 
constraint. In this case,  $R_{h.c}$,  the hard-core radius is held fixed 
at a value of 0.5 and we will consider the nature of the ground state 
structures on varying $R_1$, the range of the attractive interaction.  
Recall that while our focus is on chains of spheres (or disks), the 
tethering constraint does not play a role for the selection of 
compact conformations.   The role of the tethering constraint 
here is simply to enhance
the degeneracy of the ground state. This is because
there is an exponential (as a function of the length) 
number of ways in which one might
accomodate the tether 
through a packed configuration of beads.
A well-known example is provided by Hamiltonian walks,
which are the ensemble of configurations of chains filling a 
a finite sized square lattice\cite{vanderzande}.
For simplicity, here we will consider just
the degeneracy of the underlying conformation of spheres
without taking into account this extra degeneracy. \\

{\sl Two-dimensional case}\\

In $d=2$ it is well known that there is a unique way of packing 
hard disks so that the packing is most efficient. This optimal packing is
performed by placing the centers of the disks on a triangular lattice,
whose lattice parameter equals the disk diameter.  Each disk is then
surrounded by six nearest neighbors.  This is indeed the ground state 
structure of tethered disks when $R_1=2R_{hc}$.  
Let us first discuss the nature 
of the ground state structures for arbitrary values of 
the attraction range $R_1$ in the thermodynamic limit.
Finite size effects are less relevant here  and will be considered
carefully for the
tube and semi-flexible chains considered later on in this
Section, and,
for the two-dimesnional case, will be
briefly discussed at the end of this sub-section.

When the number of disks or spheres becomes large,
one expects the ground state to be a translationally invariant lattice.
When  $R_1 < 2R_{hc}$, the disks are unable to avail of the attractive 
interaction and one obtains a swollen phase -- all conformations of the 
chain which do not violate the hard disk constraint are equally likely. 
We now turn to an analysis of the compact conformations that one obtains 
when  $R_1 > 2R_{hc}$.\\

\begin{figure}
\centerline{\psfig{figure=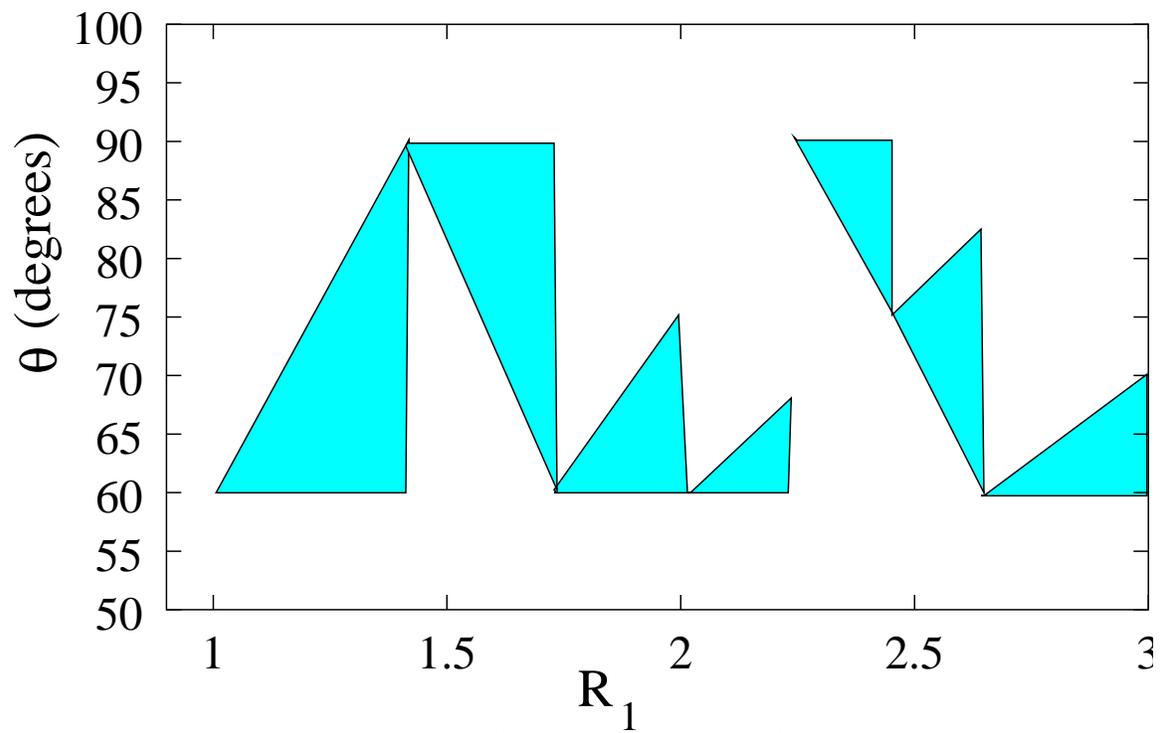,width=6.in}}
\caption{Sketch of the structural symmetry of a chain of hard disks in two
dimensions as a function of the range of the attraction (measured in units of the
disk diameter). The angle $\theta$ denotes the symmetry of the lattice.
For example $60^{\circ}$ refers to a triangular lattice and $90^{\circ}$ to
a square lattice. The shaded region shows that the ground state is degenerate with
all angles in the shaded region forming the ground state spectrum.}
\label{2dbravais_energy}
\end{figure}

Figure \ref{2dbravais_energy} shows a sketch of the winning lattice 
structures as a function of $R_1$ in the compact phase
(the lattice considered are all Bravais two-dimensional
lattices, defined by the angle between two unit vectors). The ground state is 
non-degenerate at  $R_1=2R_{hc}=1$, but on increasing $R_1$, the degeneracy goes up
(see shaded region in the figure) until a value of 
$R_1 = \sqrt{2}$, when the ground state is non-degenerate and corresponds to a
square lattice with lattice parameter equal to 1.  In this structure, both
the nearest neighbors and next-nearest-neighbors are able to avail of the 
attractive potential. \\
 
Strikingly, this pattern of non-degenerate ground states for selected
values of $R_1$ followed by segments characterized by huge degeneracies
repeats as $R_1$ increases.  Of course, in the limiting case of very large $R_1$,
one recovers the swollen phase and each disk feels an attraction with every other disk.
We have verified, with Monte-Carlo simulations, that for small systems of the order of, 
say, 25 disks, finite size effects dominate but one can be reasonably sure that 
non-translationally invariant structures do not win over the Bravais lattices 
that have been considered in our exact calculations reported in 
Figure \ref{2dbravais_energy}. \\

{\sl Three-dimensional case}\\

We have also studied the three dimensional case of a chain made of 
tethered hard spheres.  As before, when $R_1 < 2R_{h.c}$, one obtains
a swollen phase consisting of self-avoiding conformations of the chain.
When $R_1=2R_{h.c}$, one
not only obtains the fcc lattice with lattice parameter of 1 as the
ground state but there is an essentially infinite degeneracy in the
thermodynamic limit of random hexagonal close packed stackings.
This is well known in the context of hard sphere
colloids\cite{colloids1,colloids2}.
Of all these degenerate structures, the fcc structure is the only Bravais
lattice.  For the situation with $R_1 > 2R_{h.c}$, we have considered
all three dimensional Bravais lattices   and identified those that
correspond to the minima of the potential energy.  
As before,
there are huge degeneracies in the ground state conformations and
discontinuous jumps in the ground state energies as the range of the 
interaction is changed.  It is interesting to note that 
one of the ground states corresponds to stacked triangular
lattices, previously observed by Zhou et al. \cite{karplus}.  
The unit cell of this three dimensional lattice has two
triangular faces and the other faces are squares.  \\

There are several common features in our results:

1)  On increasing the range of attraction, 
one can identify an edge of compactness at which the attractive 
interaction just kicks in (when $R_1=2R_{hc}=1$) and the swollen phase
gets replaced by a triangular lattice in two dimensions and a fcc or hcp
lattice in three dimensions.  

2)  In general, the arrangement of the spheres comprising the chain
is symmetric.  A given sphere is surrounded isotropically by other spheres
to attain the best packing.

3)  One obtains a unique ground state for only certain very special
values of the range of the attractive interaction.  In general, the
degeneracy increases as one moves to the right of a plateau or as the range
of the attraction increases in magnitude.

4)  There are discontinuous jumps in the ground state energy on changing
the range of the attractive interaction.

5)  The  ground state structures in different plateaus (as in
Figure 2 for the two dimensional case) are, in general,
not the same.

In the next sub-section, we will turn to an analogous study of spheres
tethered together as before but this time with a three body constraint
that captures the tube thickness.  Note that the interacting objects
have an effective anisotropy.  Furthermore, there is an additional
length scale -- the thickness of the tube.  We will consider ground
state conformations of such a tube on varying $X$,  defined as
the dimensionless ratio of the tube thickness to the range of the
attractive interaction.  We will find many features that are
reminiscent of our results in this section: there are discontinuous
jumps in the energy on increasing $X$ with a non-degenerate ground
state at the end of each plateau.  The key difference is that because
of the inherent anisotropy of a tube, a careful relative positioning
and orientation of nearby tube segments are needed in order to avail
of the attractive interaction.  This leads to huge reductions in the
degeneracy of tube conformations.  Furthermore, for short tubes, 
at the edge of compactness, one obtains helix, hairpin and sheet
conformations which are the building blocks of protein structures.

The above picture ought to be robust on replacing the discontinuous square
well potential for the two-body interaction energy with a continuous
Lennard-Jones potential. The main difference would be the absence of
discontinuous jumps in the energy, but a plateau could still be defined by
monitoring the switching of the underlying ground state conformations, as
in the case of short stiff chains in the presence of a continuous bending
rigidity potential (see sub-section C).

\subsection{Geometry of short tubes}

In this sub-section, we will present a brief review of the results of
our simulations of short tubes\cite{Banavar,tubes}.  
The potential energy is as before but with
a three body constraint that does not allow any triplet radius to 
be smaller than the tube thickness $R_0$.  Again, we hold the 
successive tethered spheres at a mutual distance of 1 unit,
the hard core radius of the sphere $R_{h.c.}$ is held fixed at $0.55$ and
the range of the attractive interaction is maintained at 1.6 units.
We now vary the tube thickness $R_0$ (and hence the dimensionless ratio
$X$) and seek to determine the conformation(s) for which the potential
energy is as low as possible.  Thus we seek the tube analogs of the Bravais 
lattice structures that we discussed in the previous sub-section. \\

\begin{figure}
\centerline{\psfig{figure=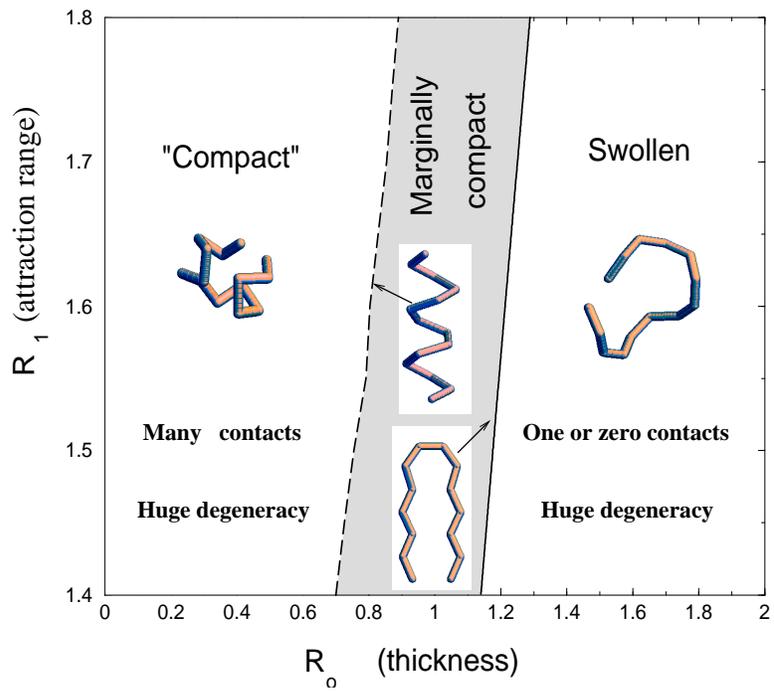,angle=270,width=6.in}}
\caption{Sketch of phases, their characteristics
and the associated ground state structures of short tubes in the $R_0$--$R_1$
plane at low temperatures.}
\label{marginal_compactness}
\end{figure}

As before, when $X$ is very  large  compared  to 1, the tube is 
so fat that it is unable to benefit  from the  attractive  
interactions.  The constraints of the three body interaction dominate 
(the pairwise interaction plays no role) and one then  obtains a 
swollen phase -- all self-avoiding conformations of the tube are equally likely
and the vast majority of them is not effective in filling the space in the core
of the structure. Note that
globular proteins fold in order to squeeze out water from the core of the 
folded state,
which houses the hydrophobic amino acids.
At the other extreme, for a tube with a  very  small  $X$  compared  to  1
(very thin tube),  
one  obtains 
many conformations, leading to an energy landscape studded with numerous multiple minima.
We thus expect a transition to occur for intermediate value of $X$, when the 
thickness is comparable to the attraction range (see Figure 
\ref{marginal_compactness}). 
We shall denote the resulting conformations near this transition as being
marginally compact structures because they
lie at the edge of the compact phase --
a  small increase in the tube thickness is sufficient to change these into
swollen conformations  unable to avail of the attractive potential.  Protein 
structures live in the marginally compact phase because the side chains
of the amino acids, on the one hand,  
determine the tube thickness (note that steric overlaps of the 
side chains has to be respected) and, on the other, control the range of the
interactions (the atoms of nearby sidechains interact through a screened, 
attractive short range interaction).  This self-tuning is a marvelous attribute
of proteins. \\

Our goal is to study the ground state conformations on varying the tube thickness.  For
a short tube, made up of $N$ balls, subject to pair-wise attractive interactions, let us
denote by $N_c (N, R_0)$ the maximum number of contacts (each with an energy of -1 
--  Eq. (\ref{two_body_potential})) that can be made 
without violating both the hard-core and the 
three body constraints.  Quite generally, one would expect that, for a fixed $N$,
$N_c$ is a decreasing function of $R_0$ and that the decrease occurs in 
discrete steps corresponding to the inability to form the same number of
contacts as before. Physically, of course, for a given contact energy (or equivalently 
the number of contacts), one would choose the largest possible thickness  in order to 
provide as much internal wiggle room (for the amino acid side chains, in
the case of a protein) within the tube as possible. \\

We have carried out Monte-Carlo simulations by using the standard simulated annealing
algorithm in order to find the ground state of the thick polymer. This basically
entails starting from a swollen disordered configuration at a high temperature,
$T$, and monitoring the 
collapse on gradually lowering the $T$.  On repeating this procedure many times,
one hopes to find a good
approximation to  the ground state conformation of the tube.
It is well known that this procedure may lead to 
the system being `trapped' in metastable minima. 
We have attempted to avoid this in two ways: first, by 
repeating the simulation many times, changing the algorithm parameters (such as the
amplitude of the dynamical moves), and 
second, by comparing the annealing performance 
with that of a recently proposed algorithm \cite{landau} for
finding the density of states of a system reliably.
The results presented here are independent of the algorithm used to
obtain them. \\

\begin{figure}
\centerline{\psfig{figure=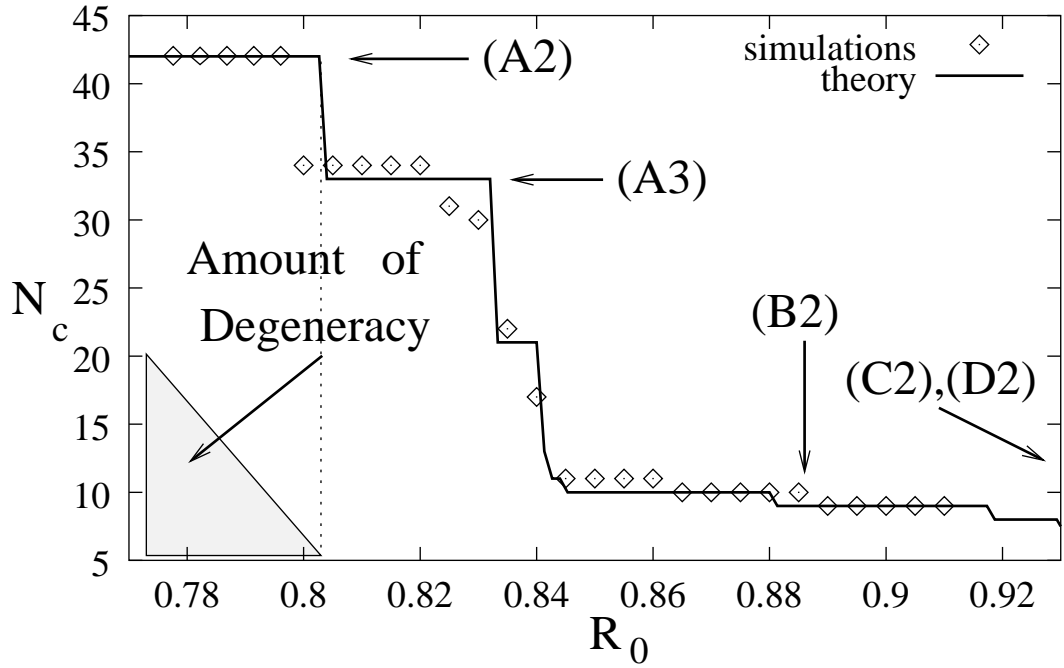,angle=270,width=6.in}}
\caption{Plot of the number of contacts, $N_c$, vs. $R_0$ for a short thick tube 
($N=14$). The labels (A2), (A3), (B2), (C2) and (D2)
refer to unique ground state structures shown in Figure \ref{configurations}.
The points show the results obtained in the simulations, whereas the line 
was obtained by means of analytic calculations carried out for idealized structures.
The triangle at the left corner depicts qualitatively the degree of 
degeneracy of the ground state structures. For example, the degeneracy
at $0.78$ is proportional to the height of the triangle at that point and the degeneracy
at $0.80$ is $1$.}
\label{NcvsR0}
\end{figure}

\begin{figure}
\centerline{\psfig{figure=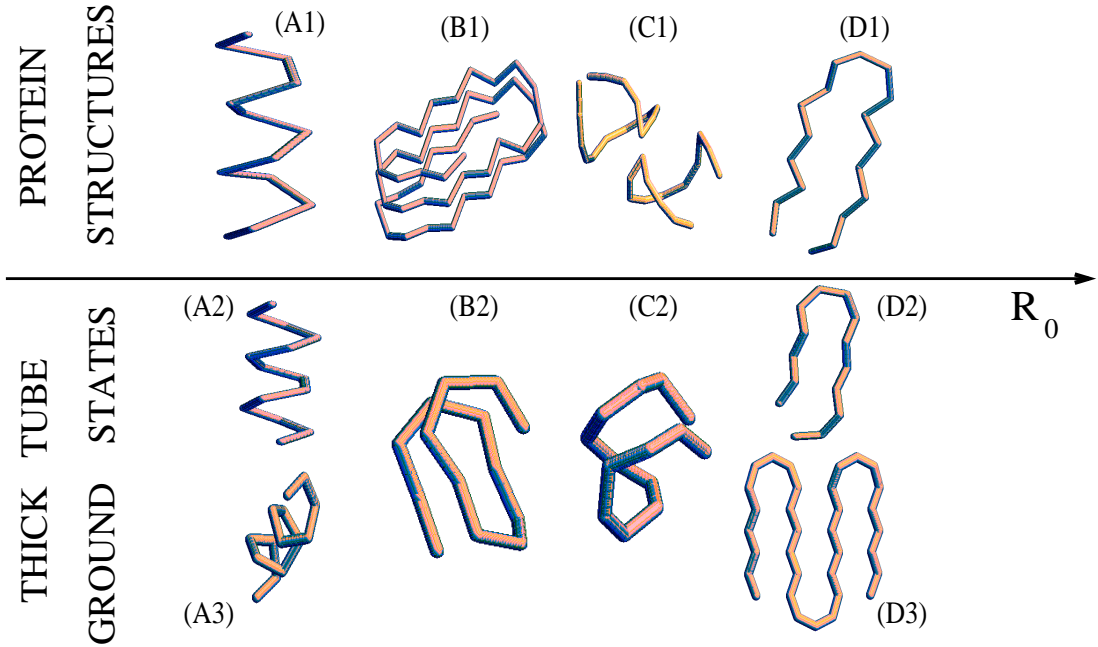,angle=270,width=6.5in}}
\caption{Building blocks of biomolecules and ground state structures associated  
with the marginally compact phase of a short tube.  The axis in the middle 
indicates the 
direction along which the tube thickness $R_0$ increases.  
The top row shows some of the building blocks of biomolecules, while the 
bottom row depicts the corresponding structures obtained as the 
ground state conformations of a short tube. 
(A1) is an $\alpha$-helix of a naturally occurring protein, while 
(A2) and (A3) are the helices obtained in our calculations -- (A2)  has 
a regular contact map and is obtained when $R_0 = 0.80 $ whereas (A3) 
($R_0 =   0.83 $) is a distorted helix 
in which the distance between successive atoms along the helical axis 
is not constant but has period $2$. (B1) is a helix of strands in the alkaline protease 
of pseudomonas aeruginosa, whereas (B2) shows the corresponding structure 
($R_0 = 0.88 $) 
obtained in our computer simulations. 
(C1) shows the ``kissing'' hairpins of RNA 
and (C2) the corresponding conformation obtained in our simulations 
with $R_0=0.95$. Finally (D1) and (D2) 
are two instances of quasi-planar hairpins. 
The first structure is from 
the same protein as before (the alkaline protease of pseudomonas aeruginosa) while 
the second is a  typical conformation found in our simulations when 
$R_0>0.98$. All the cases shown correspond to tubes of 14 spheres, 
except for 
the sheet-like structure (D3), which employed 33 spheres. This
figure also appeared in Ref. [20].   }
\label{configurations}
\end{figure}

We have carried out simulations for several values of $N$ 
and present the scenario for 
$N = 14$, which is representative of values of $N$ between $6$ and $20$.  
A schematic ``phase diagram'' is shown in Figure \ref{marginal_compactness}, 
while Figure \ref{NcvsR0} is a plot of $N_c$ 
(or equivalently the negative of the ground state energy) as a function of the tube 
thickness. For small $R_0$ one gets a highly degenerate phase  with $N_c$ saturated at 
a value of $45$ for $N=14$ -- the conformations that the tube adopts depend rather 
strongly on the details of the pair-wise potential. For $R_0$ between $\sim 0.75$ and 
$0.8$, there is an energy plateau in which the  degeneracy is greatly reduced and 
helices are the ground states. Furthermore, for the tube with the largest thickness in 
this plateau, one obtains a specific helix as the unique ground state (see upper 
conformation in ``Marginally compact phase'' of Fig. \ref{marginal_compactness} 
and Fig. 
\ref{configurations} (A2)). For $R_0$ 
between $0.8$ and $0.98$, several classes of conformations including saddles (which are
planar hairpin conformations distorted into  three dimensional structures) (see 
Fig. \ref{configurations} 
(C2)), generalized helices (in which the distance between successive balls along
the helical axis is not constant but is periodic) (see Fig. \ref{configurations}
 (A3)), helices made up of
strands (see Fig. \ref{configurations} (B2))  and other more disordered 
conformations compete.\\

At the end of each plateau in Figure \ref{NcvsR0} (there are as many as three major 
plateaus in this range of tube thickness, and each plateau comprises 
several distinct sub-steps or smaller plateaus), we find an ordered 
unique ground state (see some examples in Fig. \ref{configurations}, second row).
(The robustness of our results is underscored by the fact that helices 
and sheet-like conformations emerge as the conformations of choice
at the end of some of the major plateaus for other parametrizations of the 
potential energy of interaction as well \cite{Banavar}.)
One may show that helices are excluded from being the ground states, when the tube 
thickness exceeds $R_0^{max,hel}\sim\left (\sqrt{1+R_1^2}\right)/2\sim 0.943$ which is
obtained when two parallel straight lines (successive turns of the helix treated as 
circles with infinite radius) face each other at the bond length
$R_1$.\\ 

\begin{figure}
\centerline{\psfig{figure=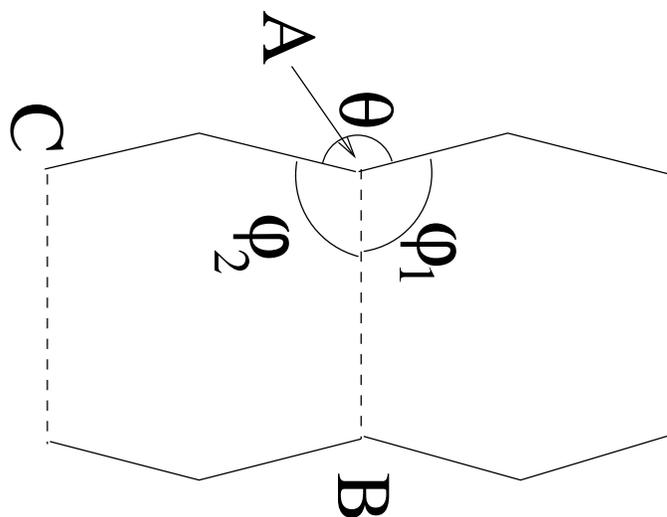,angle=270,width=3.5in}}
\caption{A schematic view of portions of two zig-zag strands facing each other.  
The angles $\theta$, $\phi_{1,2}$, and the points A, B, C which
are referred to in the text are shown.  }
\label{zig-zag}
\end{figure}

For $R_0>0.98$, the ground state structures become more and more
planar, first locally and then globally.  For large $N$, the winning
planar structures entail the combinations of strands into a sheet
structure. (We find that sheet structures persist for $N$ as large as
24, whereas the persistence length for helices is expected to be
somewhat smaller.)  For two zig-zag antiparallel strands facing each
other, one can show analytically that the maximum thickness is
obtained (leaving aside the edge effect of how the strands are
connected together in a hairpin) when one has a space-filling
conformation\cite{hull}, and furthermore when the local and the
smallest non-local radii of curvature have the same value. Indeed,
this condition leads to the following relationship between the tube
thickness $R_0$ and the interaction range, $R_1$, as defined in
eq. (\ref{two_body_potential})
\begin{equation}\label{zigzag}
R_1^2+2+\frac{R_1}{R_0}-4R_0^2=0     ,
\end{equation}
which yields a value of $R_0 \sim 1.2124$, when $R_1 =1.6$.
Eq. \ref{zigzag} can be obtained by means of simple geometrical
considerations -- one wishes to 
maximize the function ${\rm min}_{i,j,k}R_{ijk}$
with respect to $\theta$, the angle between two successive links in the
zigzags, and $x$, the distance between the two strands. 
One can readily show that the maximum thickness is obtained
for a planar configuration. In Fig. \ref{zig-zag}, the sum of the three angles 
$\theta$ and $\phi_{1,2}$ is $2\pi$. 
When the configuration is bent along the contact axis ($AB$ in Fig. 
\ref{zig-zag})
the angle between two successive links $\theta$ will be smaller and as 
a consequence the local thickness (which is $1/(2\cos({\theta/2})$) 
will be smaller. For a rotation of the left zig-zag strand 
around the axis $AC$, on the other hand,
either $\phi_1$ or $\phi_2$ will become smaller and again the thickness will
decrease.\\

The swollen phase, which occurs for even larger values of tube thickness,
has two energy plateaus.  The first of these plateaus 
has just one contact and comprises
all swollen conformations whose 
two ends make a contact (i.e. ring-like configurations)
-- the thickest tube which is able to make $1$ contact has
a unique ground state of a 
closed polygon with $N$ edges of unit length and $1$ of length $R_1$.
Likewise, the plateau of $0$ contacts 
has the limiting thickness situation of a straight, infinitely fat tube.
Indeed, starting from the zig-zag conformation, the unique 
conformations corresponding to the largest possible thickness 
compatible with a given energy (or number of contacts)
all share the intriguing property
that the local and the smallest non-local radius are exactly equal (It is
interesting to note 
that the optimal helix of Fig.  \ref{configurations} (A1) has a ratio of the smallest 
non-local to the local radius of around $0.97$ which is very close to
the corresponding value for 
$\alpha$ helices occurring in proteins \cite{Maritan,rmp}). \\

Helices and sheets are, of course, the well-known building blocks of
protein structures \cite{Pauling1,Pauling2} (see Fig. 
\ref{configurations} (A1) and (D1) for 
two examples). In addition to the prediction of
these motifs in our calculations, it is interesting to note that
some of the other marginally compact conformations bear a qualitative resemblance
to secondary folds in biopolymers.   Helices analogous to Figure 
\ref{configurations} (A3)
with an irregular contact map occur, e.g., in the HMG protein NHP6a \cite{hmg}
with pdb code 1CG7.  Fig.\ref{configurations}  
(C1) shows the ``kissing hairpins'' \cite{kiss}
of RNA (pdb code 1KIS), each of which is a distorted and twisted hairpin 
structure while Fig. \ref{configurations} 
(C2) is a saddle conformation, which is a hairpin distorted
into a three-dimensional structure.   
Figure \ref{configurations} (B1)  shows a helix of strands
found experimentally in Zinc metalloprotease \cite{zinc} (pdb code: 1KAP), whereas
Figure \ref{configurations} (B2) is the corresponding marginally compact conformation 
obtained in our calculations. \\

We point out that the results above, specific for $N=14$, apply in general for `short' 
thick polymers -- `short' here means  less 
than approximately $20-30$ monomers depending on the value of $R_0$. For longer 
tubes with thickness greater than $R_0=0.8$, the
preferred ground state 
is one in which the different portions of the tube are parallel
(see Fig.\ref{aggregate}). Indeed, this is expected to be, in the continuum, 
the ground state of a
thick tube in the thermodynamic limit. The fact that section of the tubes in contact 
must position themselves parallel with respect with each other can be rationalized 
rather simply in the continuum model. Consider  the interaction between two
cylinders the points along
whose axes are subject to a two-body potential  as  in 
Eq. \ref{two_body_potential}.  For thin tubes, 
the attraction roughly is substantially independent of
the angle between the two axis, whereas for thicker tubes there is a marked 
minimum when the two axes are parallel to each other 
(see also the discussion in \cite{tubes,seno}).
The resulting phase is similar to the nematic-like or crystalline polymer
phases obtained with semiflexible polymers and reported recently 
in Refs. \cite{muthu1,muthu2}. Our model is the first one, to our knowledge, to
predict such a crossover between secondary structural motifs 
for short polymers and these nematic structures for long chains.\\

When the protein concentration in the cell is higher than a certain critical 
threshold,  the proteins may misfold and aggregate to form amyloid 
fibers\cite{dobson}. In the case of prions \cite{PRUSINER}, for example, 
one observes a transition from $\alpha$ to $\beta$ rich structures
which favours aggregation and causes 
bovine spongiform encephalopathy (BSE) disease in cattle. The shape of
the aggregate can be visualized in a coarse grained way by thinking of different parts 
of the aggregate as portions of tubes which are aligned with respect to each other. 
Another classic example is that of DNA  condensation in the presence of 
polyvalent counterions in solution. The globule structure formed by DNA 
upon condensation is usually  a
circular donut or spool with a hole in its interior when the
DNA is short or its concentration 
is low (the stiffness of the double-stranded 
DNA molecule makes it energetically 
unfavourable to bend the tube sufficently strongly to fill the hole) 
and a `nematic-like' state with different portions of the molecule aligned 
along a local director, when the concentration is high
\cite{bloomfield}.
The tube picture can not only  explain
the observed geometries but also the crossover between 
them on changing the tube length.
If we postulate that the compact phase of a long 
tube is somewhat akin (due to the predominance of non-local contacts)
to a concentrated phase of shorter tubes, our simulations could
then qualitatively explain the experimental finding. 
Our recent simulations \cite{pre} confirm this expectation.\\

It is interesting to note that the zig-zag structures of planar configurations 
obtained for thickness $R_0\sim 1$ are not the same as found for 
$\beta-$sheets of proteins. In the latter case the zig-zag occurs perpendicular 
to the plane of the sheet. This geometry is a peculiarity of the
detailed chemistry of the backbone and of the hydrogen bond.
These peculiarities can also impact on the way proteins form amyloids
which at the present time is not precisely known
(see Ref. \cite{PERUTZ,PETKOVA}
for the most recent developments). However it is pleasing that 
the model proposed here is able to predict secondary motifs as shown in
Fig. \ref{configurations} (with the geometry of the helix exactly 
matching that of an  $\alpha$ helix) and at the same time 
amyloid-like phases for long proteins or ensembles of them.

We have previously shown that the thick polymer model defined via three
body interactions admits a continuum limit without the need to regularize the
potential. Such a continuum limit is attained by increasing the ratio
$R_0/b$, where $b$ is the distance between successive beads as in Section III.
Hairpins and zig-zags would not be found in this limit as can be
seen from Eq. \ref{zigzag} (which when the length $b$ is different from $1$
reads $(R_1/b)^2+2+\frac{R_1}{R_0}-4(R_0/b)^2=0$), 
which in the continuum limit yields
$R_0=R_1/2$ as the maximum thickness beyond which one cannot have
contacts.
When $R_0\stackrel{<}{\sim}R_1/2$ and the 
length of the polymer is sufficiently 
short, the ground state is expected to be a space-filling helix,
of the same kind as the one found in \cite{Maritan}. As the 
length increases, the ground state should cross over to a nematic-like 
structure in which different portions of the tube are straight 
and arranged on a triangular lattice when seen in a plane perpendicular to
their axes.

\subsection{Ground states of short stiff chains}

In this section we briefly study the nature of the ground states
of a semiflexible polymer chain on increasing the stiffness.
Our main aim is to understand the
similarities and differences with the case of thick tubes.
The system Hamiltonian is 
\begin{equation}
{H}=\sum_{i<j} V\left(r_{ij}\right) - 
\kappa\sum_i \vec r_{i,i+1}\cdot\vec r_{i+1,i+2}  ,
\end{equation}
where, $V$ is the
potential in Eq. \ref{two_body_potential}, characterized by
$R_1=1.6$, $R_{h.c.}=0.55$, and where $\theta_i $ is the angle between 
consecutive bonds of the chain and $\kappa$ is the bending rigidity.\\

In order to minimize the cost associated with the stiffness 
$ E_s = - \kappa \sum_{i=2}^{N-1} \cos \theta_i$, conformations
with a higher local radius of curvature are favoured, as is the
case for tubes, for which local radii less than the tube
thickness are forbidden. A key
difference, however, is that  in the semiflexible chain the
stiffness term explicitly favors certain  portions of  phase
space, whereas, in the flexible tube model, the thickness constraint
merely forbids some regions of phase space, thereby implicitly favouring
the remaining parts of phase space in a less direct way.
As a result, the stiffness term is more effective in guiding the 
semiflexible polymer
chain towards the ground state conformations, but the corresponding zoology
of the ground state structures is not as rich as in the tube model.
A crucial difference between a tube and a semiflexible chain model
is in the non-local effects inherent in a tube description.  Recall
that in addition to constraints on the local radius of curvature,
there is a strict requirement that none of the non-local triplets
has a radius less than the tube thickness.  This requirement 
leads to strong anisotropic interactions \cite{seno} 
between nearby tube segments in
the marginally compact phase and is a feature that is entirely missing
for a semiflexible chain.
Just as found for the tubes, the semiflexible polymer exhibits a curve of the
number of contacts versus stiffness (not shown) consisting of a series
of plateaus.\\
%These structures are summarized in Fig. \ref{stiff_configurations}.\\
% summary of the results is shown in Fig. \ref{stiff1}
%where the number of contacts and the total energy (absolute value)
%for ground state conformations are shown at increasing
%stiffness $\kappa$.\\

In each of the plateaus, defined by a constant number
of contacts and linearly increasing total energy, there is a unique ground
state conformation, labelled from $I$ to $V$, and shown in 
Fig. 9 (first row).
One first obtains a kind of distorted `figure-eight' hairpin ($I$),
possibly resembling the beginning of a double helix 
\cite{note1}
, for $0\leq\kappa\leq2.4$,
then a `knot' conformation ($II$), for $2.4\leq\kappa\leq4.6$, a three-rod
hairpin ($III$) for $4.6\leq\kappa\leq9.0$, and a two-rod hairpin ($IV$)
for $9.0\leq\kappa \leq10.9$. In the last plateau ($V$),
for $10.9\leq\kappa\leq\infty$, the ground state conformation is a
straight stretched rod, with $E_s=-12.0$.\\

In contrast to the tube model, where the degeneracy of ground state
conformations is reduced only at the end of each plateau, the
semiflexible chain always has a unique ground state.  In the latter
case, there is no tuning of a relevant length scale (analogous to the
tube thickness), which controls the degree to which the attractive
two-body interaction can be availed of.  On increasing the stiffness,
one does not find either a marginally compact phase or a swollen
phase.  It is evident from conformation $(IV)$ that planarity in the
semiflexible chain model arises merely from the need of having two
straight stretched rods parallel to each other, whereas it is a more
stringent geometrical requirement in the tube model.  On the other
hand, conformation $(III)$, is very similar to what we obtain for
longer chains in the tube case.  Indeed, we expect that, in the
thermodynamic limit of chains of infinite length, the ground state, in
both cases, is a nematic-like structure of long parallel rods filling
the space with hexagonal symmetry. Unlike the tube, the stiff chain
conformations (see, for example, $II$ and the conformations in the
last row of Figure 9) do not need to satisfy any non-local anisotropic
constraints. Thus, in the latter case one expects a much higher
degeneracy with a much greater number of alternative ground state
conformations consisting of planes stacked onto each other with
parallel (or antiparallel) alignment within the same plane, but not
necessarily between different planes. \\

For low $\kappa$ values, some of the simulations found conformations
with portion of helices, or resembling helices of strands, as shown in 
Fig. 9, 
but they are never the ground state for any value
of $\kappa$, and they are not as regular as in the tube case (compared e.g.
to a long straight helix). Likewise, double helix conformation were
found in the tube model but not as ground states.
 
Some preliminary results for longer chains ($N=22$) are also shown
in the last row, which  combine the main features
of conformations $(I)$ and $(II)$ for $N=14$, namely the winding
around a central axis which is in turn formed by one end of the chain. \\

In the next section, we will present mean field calculations, complemented by
Monte Carlo simulations,  of the phase diagram of a tube in the thermodynamic limit. \\

\section{Phase diagram of a tube}

Our goal, in this section, is to elucidate the phase diagram in the 
($T$ -- $R_0$) plane, where $T$ is the temperature and $R_0$ is the 
thickness of a tube subject to an attractive pair potential of a given 
fixed range.  We will do this by means of approximate analytic treatments 
which provide useful guideposts to a numerical attack on the problem.  
Taken together, these techniques provide a picture of the nature of the phases 
and the transitions between them. \\

It is useful, at the outset, to consider existing variants of the Edwards-like
model of tethered spheres to understand what qualitatively new features are
introduced in the tube picture.  A commonly studied model discussed in the
previous section is the stiff chain which starts with standard 
tethered spheres and introduces a curvature energy term in the Hamiltonian --
the cost of local bending of the chain depends on the local radius of curvature
and the sharper the local turn, the more expensive it is.  The role of temperature
in such a model is that at sufficiently high temperatures, sharp local turns are
tolerated but as the temperature is lowered, such turns become harder
to sustain and,  indeed,  one requires that the chain segments become straighter and
straighter as the temperature is lowered.  This local curvature effect may be 
captured in a potential energy of the form of Eq. \ref{bending_rigidity_term},
and leads to a schematic phase diagram in the (T-k) plane shown in Figure 
\ref{stiff}\cite{stiffness,lise}.\\

In contrast, the tube picture has a somewhat different way of
capturing local curvature effects.  For a tube of thickness $R_0$, the
local radius of curvature can be no smaller than $R_0$.  A violation
of this {\em temperature-independent} constraint would lead to a
self-intersection of the tube and that would be prohibitively
expensive.  Thus, unlike the standard approach (see paragraph above)
in which tube segments become locally straighter and straighter as the
temperature is lowered, for a tube, certain conformations entailing
tight local turns simply do not occur at any temperature.

The fact that different parts of the tube cannot intersect with each other
(the triplet radii have to be greater then $R_0$) leads to
severe constraints on the nature of the interaction between tube
segments.  As described before, especially when the tube thickness is
tuned (by hand or automatically, as in proteins) to be comparable to
the range of the self-attraction, there are sharp constraints on the
relative distance and mutual orientation of tube segments which must
be met in order for the attractive interaction to be availed of.
There are variants of the standard sphere model in polymer physics in
which dangling ends are attached to the monomers and they interact via
an anisotropic Maier-Saupe type of interaction commonly studied in the
field of liquid crystals\cite{maiersaupe}.  The tube picture may
therefore be thought of as a new way of incorporating local curvature
constraints along with a new variant of the anisotropic non-local
interactions in a chain.  Quite remarkably, these new variants result
from elegant physical considerations of a very common every day object
-- a tube -- and lead to new polymer physics and a unified explanation
for the common characteristics of protein structures. \\

We proceed first to approximate mean field treatments of two variants
of the tube model. The first is a chain of infinitesimal coins (each
with a radius $R_0$) and the second is the model with a three-body
constraint which dictates that none of the triplet radii should be
smaller than $R_0$ (see eq. (\ref{h3body}).  It is important to note
that the mean field approximation does not take into account the
tethering properly. As a result, in the case of the chain of coins
neighbouring and non-neighbouring portions of the chains are
considered to be equivalent, whereas in the second case, the non-local
effects (giving self-avoidance) are neglected.  Indeed, the formalism
used in the mean field theory is technically similar to previous
calculations with bending rigidity and liquid-crystal-like self
interactions between the polymer beads (see e.g.
\cite{muthu1,garel}). \\

Armed with the insights from the approximate
analytic analysis, we will embark on detailed Monte-Carlo simulations,
which do not entail any approximations, to deduce the phase diagram.
 
\subsection{Mean field analysis} 

{\sl The chain of coins}\\

Let us consider a chain of $N$ infinitesimally thin
coins, each having a radius $R_0$,
whose centers are 
at $\l\{\vec{r}_i\r\}_{i=1,\ldots,N}$.   The coins cannot
co-penetrate and they are subject to a pairwise purely attractive potential 
$V_{2b}(\vec r_i-\vec r_j)$ whose argument is the distance between pairs of coin 
centers. We choose an isotropic 2-body potential promoting
compaction in order to compare our results with those of
the simulations in Section VB. The anisotropy arises from the
geometrical shape of the constituents of the chain molecule.
We postulate that the centers of the coin cannot approach each other
closer than twice a given
hard core radius $R_{hc}$.  Thus one might think of the coins as 
being complemented by hard spheres whose center coincides with the coin center.
This is necessitated in our mean field treatment in order
to allow for a $\Theta$ collapse -- without the hard sphere,
the free energy minimum would always correspond to a non-zero
density and there would be no scope for 
the swollen phase. The partition function of this system 
at an inverse temperature $\beta\equiv\frac{1}{T}$ is:
\beqa\label{partition_coins}  
{\mathcal Z}=\int \prod_{i=1}^{N} d\vec{r}_i \prod_{i=1}^{N-1}\delta\l(|\vec{r}_{i+1}
-\vec{r}_i|-1\r)e^{\l(-\beta\sum_{i<j;i,j=1}^{N}V_{2b}(\vec{r}_i-\vec{r}_j)\r)}
\prod_{i<j;i,j=1}^{N}(1+f_{ij}^{(1)})\prod_{i<j;i,j=1}^{N}(1+f_{ij}^{(2)}),
\eeqa
where $f_{ij}^{(1)}$ ($f_{ij}^{(2)}$) is $-1$ if the coins (spheres) centered in 
$\vec{r}_i$ and in $\vec{r}_j$ co-penetrate, and is $0$ otherwise. We have taken the
length of a link between two successive coins equal to $1$.
We can also take $\vec r_1=\vec 0$ without loss of generality. \\

We now introduce a density field $\psi(\vec{r},\vec{t})$, where 
$\vec{t}_i\equiv\vec{r}_{i}-\vec{r}_{i-1}$, defined as follows:
\beq\label{definition_density}
\psi(\vec r,\vec t)=\sum_{i=2}^{N}\delta\l(\vec{r}-\vec{r}_i\r)\delta
\l(\vec{t}-\vec{t}_i\r).
\eeq
We proceed to expand the effective 
Hamiltonian in Eq. \ref{partition_coins} in powers
of $f_{ij}^{(1),(2)}$ as in a standard virial expansion. 
By enforcing the definition in Eq. \ref{definition_density} through the introduction 
of a conjugate density field $\hat{\psi}(\vec r,\vec t)$, and by restricting the 
virial expansion {\em in the free energy} to two particle clusters in $f_{ij}^{(1)}$ and 
up to three particle clusters in $f_{ij}^{(2)}$ (see the Appendix for more details on 
these steps), we can rewrite Eq. \ref{partition_coins} as:
\beq\label{partition_coins_2}
{\mathcal Z}=\int {\rm D}\psi\, {\rm D}\hat{\psi} \exp{\l(-{\mathcal H}(\psi,
\hat{\psi})\r)}
\eeq
where the explicit form of ${\mathcal H}$ is given in the Appendix.
From Eq. \ref{partition_coins_2} one obtains the saddle point equations
$\delta{\mathcal H}/\delta{\psi}=\delta{\mathcal H}/\delta{\hat{\psi}}=0$.
Solving these is equivalent to finding the optimal density configurations
of the system, neglecting fluctuations. 
This means that we are making use of the mean field approximation.
However an explicit solution is not possible without further 
simplifications. A very common one (used e.g. in Refs. \cite{muthu2,garel})
consists of finding the solutions of the saddle point equations which have the form
$\psi(\vec r,\vec t)\equiv \rho\phi(\vec t)$ and consequently one also has
$\hat{\psi}(\vec r,\vec t)\equiv \hat{\rho}\hat{\phi}(\vec t)$).
While $\rho$ represents the spatial density of particles, which is assumed to
be uniform, the symbol $\hat{\phi}(\vec t)$ stands for the normalized probability that
a link is oriented along the unit vector $\vec t$.

If we now use the self-consistent saddle point equations and insert them back into
Eq. \ref{partition_coins_2}, we find that the mean field free energy functional
per particle to be minimized is:
\beqa\label{mean_field_functional_coins}
f_{mf}(\rho,\psi)=\int d\vec{t} \phi(\vec t)\log{\l(\phi(\vec t)\r)}
+\frac{1}{2}\beta\int d \vec r V_{2b}(\vec r)\rho+\frac{2}{3}\rho\pi R_{hc}^3\\
\nonumber + B\rho^2
%\frac {8\pi ^2\rho ^2}{3} 
%\int_{0}^{R_{hc}} d r_1 \int_{0}^{R_{hc}} d r_2 \int_{-1}^{1} d\cos{\theta}
%r_1^2r_2^2 H(R_{hc}^2-r_1^2-r_2^2+2r_1r_2\cos{(\theta)})\\ \nonumber
+2\rho\pi R_0^3\int d\vec t\int d\vec t'
\l(1-\l(\vec t\cdot\vec t'\r)^2\r)^{\frac{1}{2}}\phi(\vec t)\phi(\vec t');
\eeqa
where the first term is the entropy contribution from the chain constraint,
the next three terms arise from the two-body potential (the first is the
attractive interaction and the other two are the repulsive terms and are
responsible for obtaining a $\Theta$ transition in our theory), whereas the
last term is the coin excluded volume mean field interaction which encapsulates
the anisotropy in the model\cite{onsager}, and where 
$B\equiv\frac {8\pi ^2\rho ^2}{3} \int_{0}^{R_{hc}} dr_1 \int_{0}^{R_{hc}} 
dr_2 \int_{-1}^{1} d\cos{\theta}r_1^2r_2^2 H(R_{hc}^2-r_1^2-r_2^2+2r_1r_2\cos{(\theta)})$
as shown in the Appendix. In the last formula $H(x)$ is $0$ 
if $x<0$ and $1$ otherwise. 
It is noteworthy  that
the qualitative form of the resulting phase diagram (see below and
Fig. \ref{phase_diagram_mean_field}), does not depend  on the numerical
value of $B$, but just on its presence and on it being greater than $0$
in order to render the density of the globule finite when the
system is in the compact phase.  \\

We now have to minimize Eq. 
\ref{mean_field_functional_coins} with respect to $\rho$ and $\phi$.
We define $\int d\vec r V_{2b}(\vec r)\equiv -V_0$, which is the only feature of the
pair-wise attractive potential which affects the mean field solution.
When $T$ is high ($\beta$ is small), the minimum of $f_{mf}$ occurs for 
$\rho=\psi(\vec t)=0$: this is the isotropic, swollen phase.
At a critical value of the temperature given by
$\beta_{\Theta}=T_{\Theta}^{-1}=\frac{\frac{\pi}{2}R_0^3+\frac{4}{3}\pi R_{hc}^3}{V_0}$, 
there is a phase transition into a phase with $\rho\ne 0$ and $\phi=0$, which is
physically an isotropic, globular phase. This transition is second order within the
mean field approximation ($\rho\propto(\beta-\beta_{\Theta})$ as 
$\beta\to\beta_{\Theta}^{+}$). If $T$ is further 
lowered, we find a second phase transition to an anisotropic, globule 
phase (with non-zero
orientational order parameter $\phi$) at a temperature:
\beq\label{foldingt}
T=T_F=\frac{T_{\Theta}}{1+\frac{B T_{\Theta}}{R_0^3 V_0}}
\eeq

Eq. \ref{foldingt} was derived by mapping our theory
onto the liquid crystal theory discussed in Ref. \cite{lekkerkerker}
(see Section II of that work and in particular Eqs. 18 and 27).
Another possible way to obtain the phase boundary is via the use
of Onsager's trial functions~\cite{onsager}.
(A similar procedure can also be carried out for the semiflexible 
chain model.) 
This is a first order transition since the order parameter, which is
the average of the second Legendre polynomial with
respect to $\phi(\vec t)$, displays a jump at this transition.
This transition is qualitatively similar to the
transition found in Ref. \cite{frenkel2} with a fluid of hard discs.
Note that the ratio between the two transition  temperatures, 
$T_F/T_{\Theta}$ (see Eq. \ref{foldingt}) approaches $1$ for large $R_0$ as 
$1-\frac{a}{R_0^3}$, with $a>0$ a constant.\\

The precise form of the phase diagram is shown in 
Figure \ref{phase_diagram_mean_field}, top panel. 
A qualitatively similar phase diagram is to be expected
if the chain of coins is replaced by a chain made up of 
objects with an inherent asymmetry in their geometry, such as
e.g. a cylinder.
There are two distinct regimes as a function of tube thickness: 
the first is one in which the
tube is 
`thin', which occurs when $R_0<1.5$ roughly, 
and in which there are two  
transitions as $T$ decreases, the first (the usual $\Theta$
transition) from the coil to the globule phase, 
and the second, which we shall call  a `folding
transition', between an isotropic globule and a 
compact configuration in which there is 
orientational order. In this low temperature phase, the ground state of
the chain of coins resembles the lamellar semi-crystalline polymer phase
and the  ground state of the long thick polymer (Figure \ref{aggregate}) 
discussed in the 
previous section.  

The second regime is when the thickness $R_0$ is larger than
$1.5$: we can think of this as the `thick tube' case. In this second regime
there is practically only one transition, i.e. the temperature range
over which one observes the isotropic, globule phase is vanishingly
small.  Indeed, we expect that the very existence of this phase for
thick tubes is an artifact of our mean field approximation.  
A similar situation was observed in References 
\cite{stiffness,garel} in which two transitions were predicted in the
mean field approximation for stiff polymers whereas simulations indicate that
the true solution should display a single first order transition.
There is evidence from experiments and recent theoretical results 
on the thermodynamics of proteins \cite{prl,experiment} that they may undergo 
either only one single folding transition or a $\theta$- collapse followed by 
a distinct folding transition at a lower temperature. This suggests 
that both cases, corresponding to thin and thick tubes, 
are found in protein thermodynamics.
We conclude with the observation that the absence, in our analysis, 
of a swollen phase
for very thick tubes at zero temperature, contrary to expectations based on common 
sense considerations that the attractive interactions cannot be 
availed of when the tube is too thick,  is an 
artifact of the mean field approximation. 
 \\

We now briefly compare the phase diagram resulting from the
mean field approximation in our model with that in Fig. 10,
stemming from the corresponding approximation in the stiff polymer model.
The three main phases present in both cases are the same.
There is however a notable difference in the phase boundary shape. 
For the tube, there is a transition at low 
$T$ between the anisotropic globule and swollen phases on 
inflating the tube. In contrast, for a stiff chain,
unless the bending rigidity penalty is 
infinite at zero temperature one is always in the compact phase.
Furthermore,  the collapse of stiffer chains occurs at higher $T$,
while that of thicker tubes occur at lower $T$. \\

{{\sl The tube with a three-body radius constraint}}\\

We briefly give the results of the same mean field treatment applied to a
tube-like polymer with a three body constraint (see preceding section).
Here once more the suitable order parameters are $\rho$, the density, 
and $\phi(\vec t)$,
the link orientational order parameter.   We have also considered 
the presence of a standard two-body hard core which prevents the 
transition temperature from being infinity, when one has a 
vanishingly thin tube i.e. when $R_0=0$.                  
The phase diagram is shown schematically 
in Figure \ref{phase_diagram_mean_field}, bottom panel.
Unlike the chain of coins, there is no region in which there are two
transitions. The transition temperature is $T_{\Theta}=
\frac{V_0}{2B(R_0)+\frac{8}{3}\pi 
R_{hc}^3}$, where $B(R_0)\equiv\frac{1}{V}\int\int\int d\vec r_1 d\vec r_2 
d\vec r_3 f_3(\vec r_1,\vec r_2,\vec r_3)\delta(|\vec r_3-\vec r_2|-1)$,
which goes to $\infty$ as $R_0\to\infty$. 
In this formula $f_3$ is $0$ if the radius of the circle 
constructed with the triplet
$\vec r_{1,2,3}$ is greater than $R_0$ and $-1$ otherwise: this term
acts as an effective thickness-dependent two-body hard core.  $V$ is the
volume occupied by the system in the canonical ensemble.  
The $\Theta$ collapse is from a swollen phase to an isotropic globule
if $R_0<\frac{1}{2}$, and to an anisotropic globule otherwise.
There is a transition from the thin tube to the thick tube limit at low
temperatures as the thickness increased past $R_0=1/2$. 
Unlike the chain of coins, the transition is second order in both
cases. It is interesting to note that that the
worm-like chain model\cite{siggia}, eq. (9), subject to a potential, which
promotes compaction, 
would yield a qualitatively similar phase 
diagram in the mean field approximation. This is so because,  in first order 
in the cluster expansion, 
the tube is important only in disallowing strong bends as happens also for 
polymers with stiffness. 
This also explains why the
boundary between thin and thick tubes is located at $R_0=1/2$: 
the local constraint on the bending angle imposed by the thickness
is given by $\vec R_{i,i+1}\cdot\vec R_{i+1,i+2}\ge {1-1/(2R_0^2)}$, and 
there is no restriction for $R_0<1/2$. 
Indeed,
the configuration of two
neighbouring links being exactly superimposed corresponds to a local
thickness which is $1/2$. 
However, higher order terms in the virial expansion,
that have not been considered here, would begin to capture the
non-local thickness effects when different portions of the tube come
into contact with one another. This results in differences
between the tube constraint and 
the worm-like chain model and also alters the
threshold value between the thin and thick tube limits. 

Even though the local constraint in our mean field is not the full
story and does not properly begin to capture the self-avoidance
imposed by non-local constraints, it already allows one to identify an
important difference between our model and the worm-like chain when
the continuum limit is approached.  From the inequality $\vec
R_{i,i+1}\cdot\vec R_{i+1,i+2}\ge {1-1/(2R_0^2)}$, one can infer after
some some standard algebra that, in the swollen phase, the persistence
length of a tube of thickness $R_0$ scales as $R_0^2/b$ as $b\to 0$
(see Ref.\cite{capsid} for a calculation of the persistence length of
a thick polymer applied to the case of double-stranded DNA molecule).
This is connected to the fact 
that the local tangent to the tube axis is continuous and
differentiable and thus the axis (centerline) of the continuum tube is
twice differentiable, whereas the curve corresponding to the continuum
limit of a semiflexible polymer is only differentiable once.  This
means that the thick polymer model can be of use in cases in which all
three Frenet vectors need to be defined in the continuum limit.

The mean field calculations provide us with useful insight on
what kinds of phases one might expect, the possibility of qualitatively
distinct behaviors of thin and thick tubes and the possibility in some cases
for two transitions on lowering the temperature starting from the high 
temperature swollen phase.  Armed with this information, we now proceed to
careful Monte Carlo simulations, which are not subject to the
approximations used in the mean field calculations.  \\

\subsection{Monte-Carlo evaluation of the phase diagram}

We now consider a tube of a given thickness, 
schematized as explained above with
a three body effective constraint. 
The methods employed are Monte-Carlo simulations, with the parallel tempering 
or multiple Markov chain technique (see \cite{orlandini}). 
A number of replicas (from $11$ to $16$ in our simulations)
of the polymers are equilibrated simultaneously at different 
temperatures, allowing the possibility of two replica neighbours in $T$ to
exchange their configurations after a fixed number of steps.
In all formulas below, $\langle \cdot\rangle$ denotes ensemble averaging. 
For each thickness considered, we calculate the specific heat $C$ using
\beq\label{specificheat}
C(T,R_0)=\beta^2 \l(\langle E^2\rangle-\langle E\rangle^2\r),
\eeq
where $E$ is the internal energy (contact number) and $\beta\equiv\frac{1}{T}$
is the inverse temperature.  The  radius of gyration $R_g$ is defined by
\beq\label{girationradius}
R_g(T,R_0)=\langle\frac{\sum_{i=1}^{N}
\l(\vec r_i-\vec r_{cm}\r)^2}{N} \rangle^{1/2},
\eeq
where $\l\{\vec r_i\r\}_{i=1,\ldots,N}$ represent the coordinates of the 
$N$ beads and $\l\{\vec r_{cm}\r\}$ the coordinates
of the center of mass of the tube. \\ 

We also record three probability distributions:  the distribution of the
chirality $\chi$, which is a number between $-1$ and $1$
defined for every consecutive quadruplet of beads along 
the chain by
\beq\label{chirality}
\chi\equiv \vec{r}_{i,i+1}\cdot\l(\vec{r}_{i+1,i+2}
\times\vec r_{i+2,i+3}\r);
\eeq
the distibution of $\cos({\theta})$, where $\theta$ is the angle between 
consecutive links along the chain; and that of $\cos({\xi})$ 
where $\xi$ is the angle
between two links which are in contact. Note that the distance between 
successive beads is held fixed at unity and  two links ($\vec r_{i+1}
-\vec r_i$ and $\vec r_{j+1}-\vec r_j$) are defined 
to be in contact when $|\vec r_i-\vec r_j|<R_1$.
We label these probability distributions $P(\chi)$, $P\l(\cos{(\theta)}\r)$ and 
$P\l(\cos{(\xi)}\r)$ respectively. For comparison, it is useful to note the
shapes of these distributions for a non-interacting random walk:
\beqa\label{random_distributions}
P(\chi) & \propto & {\rm arcsin}\left({\sqrt{1-\chi^2}}\right)\\
P\l(\cos{(\theta)}\r)\equiv P\l(\cos{(\xi)}\r) & \equiv & {\rm const.} \nonumber
\eeqa
\\

As with all Monte Carlo algorithms, finding canonical averages and
particularly the specific heat at low temperatures is difficult. The 
multiple Markov chain (or parallel tempering)  algorithm helps
alleviate this difficulty by enhancing the mobility of the chain at low $T$.
To further improve the
results at low $T$, we first performed one or more parallel tempering runs 
collecting data for the canonical averages, the specific heat and the low $T$ 
configurations, starting from an open initial condition (more or less 
straight chains are chosen for all the replicas). 
Then another run is performed with the 
replicas starting from a `folded' conformation, the one with the minimum energy found 
in the preceding runs. The latter approach allows one to 
sample more accurately the low $T$ 
configurations, giving a bias towards the `correct' ground state at very low $T$.
We have verified that the high $T$ behaviour is the same for all runs. \\

Figure 
\ref{phase_diagram_mc} shows the phase diagram in the $(T,R_0)$ plane and confirms
the mean field prediction of two quite distinct regimes, one for thin tubes
and the other for thick tubes with the boundary separating them being 
$R_0 \sim 0.8$ for the parameters used in the simulation. 
Figures \ref{gir_07} and \ref{gir_095}  show the behaviour of the 
specific heat and of the radius of gyration for two different thickness values, 
typical
of thin ($R_0=0.7$) and thick ($R_0=0.95$) tubes.  Figures 
\ref{chi_07}-\ref{xi_095} 
show typical histograms for $P(\chi), P(\cos{(\theta)}),
P(\cos{(\xi)})$. We considered polymers with $N=20,41$ and $60$ and performed
Monte-Carlo runs scanning the thickness $R_0$ in steps of  $0.05$ units. \\

{\em Regime 1: Thin tubes}

In many respects, the thermodynamic behavior of thin tubes is similar
to that reported  in
Ref. \cite{karplus}, which was meant to describe conventional polymers of
zero thickness.  Here we will describe the behavior of a tube of thickness
$R_0=0.7$ as representative of the thin tube regime.
Figure \ref{gir_07} shows a scaling plot of the radius of gyration versus
temperature.  Note the nice intersection of the three curves at 
$T_{\Theta}=1.5$ with the standard scaling of $R_g\sim N^{1/2}$ 
based on the expectation of $\nu=\nu_{\theta}=1/2$.
The specific heat however displays a peak only for $T$ approximately
equal to $1$ (see the somewhat noisy Figure \ref{gir_07} ), 
a value significantly smaller than the previous 
one. The data are suggestive of another transition 
at a lower temperature. \\

The $P(\chi)$ histogram is more strongly peaked around $\chi=0$ than one would
expect for the random walk case in the swollen phase.  In the isotropic globule
phase, along with the drop in the radius of gyration, one observes a somewhat
flatter distribution, which persists down to the  lower temperature
peak in the specific 
heat at which point a multi-peaked histogram of $\chi$ is obtained.
The analysis of the distribution
of the cosine of the 
angle $\theta$ between successive links suggests that,  at high $T$ (in the swollen
phase) it is almost flat as expected from random walk considerations.
% except for a depletion
%around $cos (\theta) = 1$ due to the bending constraint imposed by the tube
%thickness.
In the isotropic globular phase,  values of $\cos{(\theta)}$ 
corresponding  to local tube segments
which are not tightly bent are penalized;  at the lowest temperature shown,
this effect is more pronounced (and two peaks develop for small values of
$\cos{(\theta)}$).
Strikingly, at temperatures below the `$\Theta$-like' transition, the
values of $\cos(\theta)$ which are not penalized correspond to a local
thickness roughly between $0.7$ (the constraint imposed by the tube)
and slighly more than $0.8$ -- it is as if 
`locally thin' tubes
only dominate the thick polymer distribution probability after the $\Theta$-like 
collapse. 
It is interesting to note that, with $R_1=1.6$, in order for beads
$i$ and $i+2$ to make contacts, it is necessary that
$R_0<0.83\ldots$, which is
close to the thick threshold size. 
%In other
%words, after the theta collapse the chain will restrict its configurations only
%to those whoch allow itself to pair all beads $i$ and $i+2$.
Finally, the $P(\cos{(\xi)})$ histogram shows that
the lower temperature peak in the specific heat
is associated with a mild increase in the probability
of contacts to occur between anti-parallel  and parallel links.  \\

In Ref. \cite{karplus} it was suggested that, 
in a model similar to ours without the tube constraint, 
on lowering the temperature,
first, there is a $\Theta$
transition from the coil to the globule phase, 
equivalent to the usual gas-to-liquid transition.
This is signalled by a drop in the radius of gyration as well as by a shoulder in
the specific heat.
Subsequently, the globule crystallizes (as in a liquid-to-solid transition
for clusters of hard spheres) giving rise to the first peak in the specific heat.
This is a first order transition in contrast to the second-order
transition of the standard $\Theta$ collapse.  Finally,  there is a
third transition, which the authors refer to as a solid-to-solid first order
transition. 
Exact enumerations \cite{flavio} of interacting 
self-avoiding walks on a two-dimensional lattice suggest a similar behavior
with a drop in the radius of gyration at a higher $T$ than the one at 
which the specific heat displays a peak, and at
a yet lower $T$ one finds a second specific heat peak. \\
 
It is our belief that the thermodynamics of a thin tube is similar
to that proposed in Ref. \cite{karplus} up to the first specific heat peak.
At lower temperatures, the cylindrical shape of the tube leads to different
behavior.  Our numerical evidence is somewhat weak on whether the second specific
heat peak corresponds to a physical transition.  In any case, we expect that
such a transition, if it existed,  would be different from the solid-to-solid
transition postulated for polymers. \\

{\em Regime 2: Thick tubes} 

\begin{figure}
\centerline{\psfig{figure=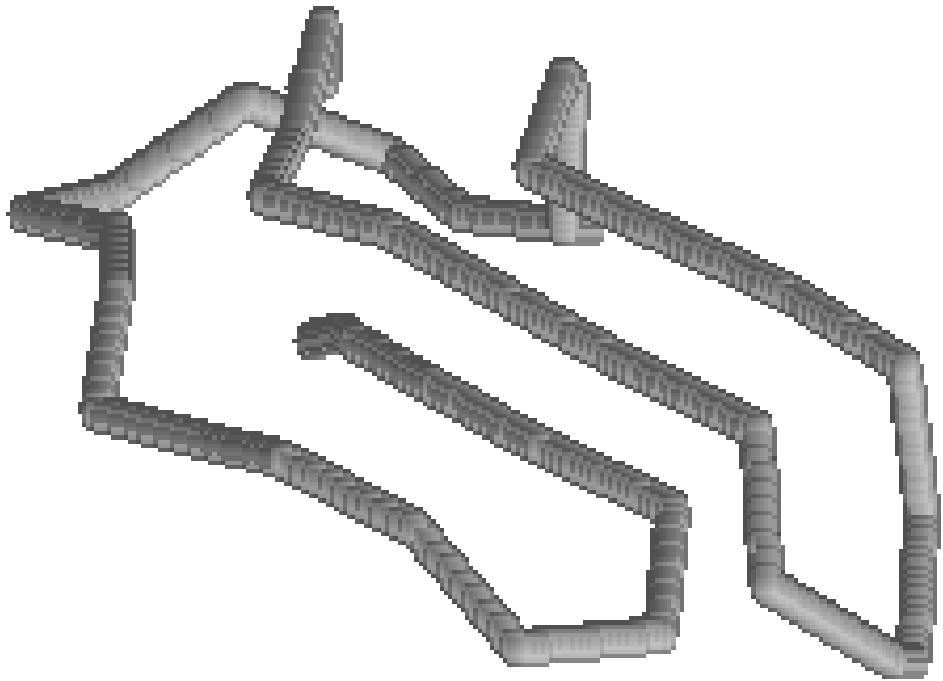,angle=270,width=6.5in}}
\caption{One of the low energy states for a tube
of thickness $0.9$ with $41$ beads. Note the parallel alignment of
neighbouring tube segments.}
\label{aggregate}
\end{figure}

For larger thicknesses (roughly from $R_0=0.8$ upwards), the
thermodynamic behaviour is different.  In this case, the point in
which the scaled curves for $R_g$ intersect and the location of the
peak in the specific heat (see Figure \ref{gir_095}) are not very
different as it was in the case of thin polymers. This suggests that
in this case there is a single transition. Also the data for the
specific heat per monomer, within the Monte-Carlo errors, suggest that
this quantity diverges at the transition  linearly with the chain
length. We thus predict that this is a first order direct transition
from a swollen phase to a `nematic-like' globular phase. This
prediction is in agreement with the mean field analysis of the
preceding subsection and is also confirmed by other data (see below).
That the low temperature phase is indeed characterized by segments of
the tubes trying to position themselves parallel or anti parallel with
respect to each other is clear from the distribution $P(\cos{\xi})$,
which, at low $T$, has sharp peaks around $-1$ and $1$.  As is the
case for the continuum ground state of long tubes, one may conjecture
that for long tubes the ground state will have tube segments aligned
with respect to one another similar to the Abrikosov flux lattice, as
in Figure \ref{aggregate}.  It should be noted that the segments of
the tubes which face each other are more or less straight only for
tubes that are not too thick (around $R_0=0.9$ in our simulations, see
Figure \ref{aggregate}).  For larger tube thickness, the thickness
constraint leads to an absence of a peak in the distribution
$P(\cos{(\theta)})$ around 1 (See Figure \ref{theta_095}).  The
analysis of $P(\chi)$ (Figure \ref{chi_095}) reveals that at high $T$
the chirality distribution is wedge-shaped, so the peak at $\chi=0$ is
more enhanced with respect to the case of low thickness. This shape
persists almost up to the transition point, and there are no
pre-transition effects as in the thin polymer case, consistent with a
sharp first order transition. Below the transition, three moderate
peaks develop, one for $\chi=0$, which is not present for $R_0=0.7$,
and the other two roughly symmetrically at the edge of the $\chi$
spectrum visited by the system.  The central one likely originates
from planar or quasi-planar structures in the sampled configurations
-- a planar structure has $\chi=0$. \\

The scaling collapse of the radius of gyration data are consistent with
both $\nu=\frac{1}{2}$ as in the standard $\Theta$ point
transition and $\nu=0.588\ldots$ as would pertain to a first order   
transition with coexistence at the transition point (the exponent of
polymers in the swollen phase). It is interesting to note that at the
critical value of $R_0$ at zero $T$ (precisely $R_0 = 1.2124\ldots$),
one obtains, as the ground state, compact configurations in $d=2$ whose
$\nu=\frac{1}{2}$.\\

Strikingly, the  `triple point' in the phase diagram (Figure \ref{phase_diagram_mc}),
which separates the `thin' and `thick' tube regimes (corresponding to a
tube thickness of around $0.8$) 
coincides with the thickness corresponding to the
onset of the marginally compact region in the minima analysis for short polymers
detailed in Section III. It is instructive to compare the phase diagram in Figure
\ref{phase_diagram_mc} with that 
of a stiff polymer (Figure \ref{stiff})
with no thickness (see e.g. Ref. \cite{stiffness,lise,garel}). 
Stiff polymers are known to undergo either two transitions, one second and the other
first order, or only one first order transition, according to whether the stiffness is small
or high. The thickness of our tube prevents  it from having sharp bends and this is
reflected in the tube having a bigger persistence length than a polymer with no
thickness: in other words, the thickness acts {\em locally} as a stiffness.
However, the thickness also has an important non-local role 
both in providing a simple
mechanism through which distant portions of the polymers must orient 
themselves selectively
in order to take advantage of attractive interactions, and, more important, 
in providing  a natural
way in which secondary motifs such as helices and sheets
arise with no need of heterogeneity in the model.
As we discussed above,  no such secondary structures appear in the resulting ground states 
of  stiff polymers, whereas with thick polymers we obtain both 
$\alpha$ helices and $\beta$ sheets. 
Also, thick discrete polymers display a planar phase at high 
values of the thickness which does not happen for stiff polymers. \\

We have also performed an analysis of the `unfolded state', i.e. 
of the {\em ensemble} of
tube configurations in the swollen phase at temperatures just above the
transition.
Again the situation is different for thin
tubes and thick tubes. In the former 
case, the unfolded state is one in which some contacts are already formed, 
though with no regularity typical of the crystalline phase. In the latter case, on 
the other hand, consistent with the fact that the transition is found to be 
first order, we 
observe that just above the transition point the chain is still indistinguishable from 
a typical configuration in the swollen phase. This is supported by a study of the
$P(\cos{\xi})$ distribution --
for a thick tube, there are no precursors to the transition (see Figure \ref{xi_095}).
This confirms, as is observed in small proteins, that the
folding transition is sharp or an all-or-none kind. 

{\it Regime 3: Very thick tubes}

We can finally identify, in our Monte Carlo simulations, a regime not found in the mean
field calculations of a swollen phase corresponding to very thick tubes.
As expected from physical considerations, very thick tubes ($R_0>1.2124\ldots$)
are unable to avail of the attractive interactions and one obtains a swollen phase at 
all temperatures. This is in contrast to our mean field result within which the 
threshold thickness was essentially infinity.  Note that such a swollen phase 
does not occur for chains with a bending energy constraint. \\

\section{Conclusions}

In summary, we have introduced the concept of a thick polymer and have
studied the phase behaviour of such tubes. A key point to describing
tubes is the need to have a three body constraint for ensuring that
the local radius of curvature is not smaller than the tube thickness
and that the tube does not self-intersect. Such a description deftly
avoids the need for a singular interaction energy in the continuum
limit as is the case with the classic Edwards model. There are three
regimes of tube thicknesses (measured as a ratio of the thickness to
the range of the self-attraction) with distinct phase behaviours. 
Generally, the categorization of a given polymer
chain, in terms of the corresponding thickness regime, will crucially
depend on its specific stereochemical properties, and the tube
thickness may well not be a relevant physical parameter in some
cases. On the other hand, there are cases, such as proteins, where the
presence of bulky side chains attached to the main backbone naturally
calls for a tubelike description. Indeed, the intermediate thickness
regime leads to space-filling marginally compact conformations and for
short tubes the ground state structures are helices and planar sheets
with the same geometry as in secondary motifs in proteins. Moreover, this
phase has many advantages associated with it which are exploited by
nature housing biomolecular structures in it. The tube picture
provides a natural way of connecting conventional polymer phases with
the biomolecular phase. 

We have also compared the ground states and the thermodynamics of 
a short tube with those of a chain of spheres with a non-zero
stiffness (worm-like chain model in its continuum version).  No
secondary structures and no planar ground state structures appear, in
general, for the stiff polymer model, as demonstrated in our
computational analysis of short chains.  Also, while stiff polymers
can undergo two different transitions on varying the temperature for
certain values of the bending rigidity parameter, they do not exhibit
a swollen phase at zero temperature as does a fat tube. Furthermore,
thick polymers not only provide an elegant explanation for the novel
phase selected by Nature to house biomolecular structures but also
provide a smooth link between this phase and conventional polymer
phases.  More fundamentally, the tube picture incorporates the
anisotropy inherent in any chain model and provides the simplest
geometrical construction that captures the essential features of any
chain molecule.

On increasing the length of the polymer, or the number
of polymer chains,  there is  some similarity
between the tube picture and the worm-like chain model. When the
tube is longer, we observe, in computer simulations, a crossover to
semi-crystalline structures with different portions of the backbone
chain lying parallel to one another. Our mean field analaysis
yields a phase diagram in the thermodynamic limit which is
somewhat similar to the stiff chain phase diagram (see
Fig. \ref{stiff},\ref{phase_diagram_mean_field}). There are clear
differences as well arising from the fact that local
distortions with a radius of curvature less than the tube thickness
are disallowed for a flexible tube, whereas, for a stiff chain, there is an increasing
penalty as the temperature is lowered and the local bending increases.
Moreover, and most significantly, the connection with the biomolecular
phase present in the phase diagram of short tubes allows us to
rationalize the ubiquitous formation of amyloid fibrils in misfolded
proteins as a biopolymer crystalline phase. \\

The simplified thick
polymer model presented here  does not, of itself, lead  to
tertiary arrangement of elements of local secondary 
structure.  We have recently shown \cite{pnas,pre}
that this likely arises from the geometrical constraints imposed by hydrogen
bonds. \\

Another intriguing direction for future research is to consider the role of
amino-acid heterogeneity in the two body potential, which is well
known to play an important role in determining the folded structure of
a polypeptide (see e.g. Ref.\cite{sauer} and references therein).  An
extension of ideas presented here can be used to describe surfaces of
non-zero thickness. Such a description entails the use of $4$-body
potentials \cite{BM}. An exploration of the phase diagram of such
systems, which are relevant to membranes and surfaces, is likely to
yield novel and rich behaviour as well.\\

{\it Acknowledgments:} We are indebted to Trinh Hoang, Sanat Kumar and 
Flavio Seno for stimulating discussions.
This work was supported by MURST, cofin2003, NASA, NSF
(DGE-9987589) and the Penn State MRSEC
under NSF grant DMR-0080019.

\appendix

\section{Mean field calculations}

In this Appendix we derive the formulas for the mean field free energy Eqs.  
\ref{partition_coins_2} and \ref{mean_field_functional_coins}. 
We consider only the case of the chain of coins, the case of the polymer with triplet  
constraint is similar. The symbols are as in Section V A.
We start from the partition function in Eq. \ref{partition_coins}:   
 
\beqa\label{a1} 
{\mathcal Z}=\int D\psi\int \prod_{i=1}^N d\vec r_i
\int \prod_{i=1}^{N-1} d\vec t_i 
\delta\l(|\vec{r}_{i+1}-\vec{r}_i|-1\r)\\ \nonumber 
\delta\l(\vec{t}_i-\vec{r}_{i+1}+\vec{r}_i\r)
\delta\l(\psi-\sum_{i=1}^N 
\delta(\vec r-\vec r_i)\delta(\vec t-\vec t_i)\r)\\ \nonumber 
e^{\l(-\beta\sum_{i<j;i,j=1}^{N}V_{2b}(\vec{r}_i-\vec{r}_j)\r)} 
\prod_{i<j;i,j=1}^{N}(1+f_{ij}^{(1)})\prod_{i<j;i,j=1}^{N}(1+f_{ij}^{(2)}). 
\eeqa 
where, as in the text (Section V A),
$f_{ij}^{(1)}$ ($f_{ij}^{(2)}$) is $-1$ if the coins (spheres) centered in 
$\vec{r}_i$ and in $\vec{r}_j$ co-penetrate, and is $0$ otherwise. 
We can rewrite Eq. \ref{a1} as: 
\beqa\label{a2} 
{\mathcal Z}=\int D\psi e^{-\beta\frac{1}{2}\int{d\vec r_1 d\vec r_2 
\psi(\vec r_1,\vec t_1)V_{2b}(\vec{r}_1-\vec{r}_2)\psi(\vec r_2,\vec t)}}
\int \prod_{i=1}^N d\vec r_i
\int \prod_{i=1}^{N-1} d\vec t_i 
\delta\l(\vec{t}_i-\vec{r}_{i+1}+\vec{r}_i\r)\\ \nonumber
\delta\l(|\vec{r}_{i+1} 
-\vec{r}_i|-1\r) \delta\l(\psi-\sum_{i=1}^N  
\delta(\vec r-\vec r_i)\delta(\vec t-\vec t_i)\r) 
\langle{\mathcal Z}_{int}\rangle(\psi). 
\eeqa 
We have called: 
\beqa\label{a3} 
\langle{\mathcal Z}_{int}\rangle(\psi)= 
\frac{\int_{\psi}\prod_i d\vec r_i d\vec t_i 
\prod_{i<j;i,j=1}^{N}(1+f_{ij}^{(1)})\prod_{i<j;i,j=1}^{N}(1+f_{ij}^{(2)}) } 
{\int_{\psi}\prod_i d\vec r_i d\vec t_i}, 
\eeqa 
where with the symbol $\int_{\psi}$ we mean integration with respect to 
the density $\psi(\vec r,\vec t)$: 
\beqa 
{\int_{\psi}\prod_i d\vec r_i d\vec t_i}\equiv 
{\int\prod_i d\vec r_i d\vec t_i} \delta\l(|\vec{r}_{i+1} 
-\vec{r}_i|-1\r) \delta\l(\vec{t}_i-\vec{r}_{i+1}+\vec{r}_i\r)
\\ \nonumber \delta\l(\psi-\sum_{i=1}^N  
\delta(\vec r-\vec r_i)\delta(\vec t-\vec t_i)\r). 
\eeqa 
Through a standard virial or cluster expansion\cite{lekkerkerker} 
truncated at second order in $f^{(1)}$ and at first order in 
$f^{(2)}$\cite{note2}, 
we get: 
\begin{center}
\beqa\label{a4} 
\log{\l[\langle{\mathcal Z}_{int}\rangle(\psi)\r]}\sim  
+  \frac{1}{2V}\int d\vec r_1{\int d\vec r_2{\psi(\vec r_1)\psi(\vec r_2) 
f^{(1)}(\vec r_1,\vec r_2)}} \\ \nonumber 
 +  \frac{1}{3V}\int d\vec r_1{\int d\vec r_2{\int d\vec r_3{\psi(\vec r_1) 
\psi(\vec r_2)\psi(\vec r_3)f^{(1)}(\vec r_1,\vec r_2)f^{(1)}(\vec r_2,\vec r_3)f^{(1)} 
(\vec r_1,\vec r_3)}}}\\ \nonumber 
 +  \frac{1}{2V}\int d\vec r_1{\int d\vec r_2{\psi(\vec r_1)
\psi(\vec r_2)  f^{(2)}(\vec r_1,\vec r_2)}},  
\eeqa  
\end{center}
where by $\psi(\vec r)$ we mean $\int d\vec t\psi(\vec r,\vec t)$. 
Now in Eq. \ref{a2} we use the well-known identity: 
\beqa\label{a5} 
\delta\l(\psi-\sum_{i=1}^N  
\delta(\vec r-\vec r_i)\delta(\vec t-\vec t_i)\r)=\\ \nonumber 
\int D\hat{\psi} e^{i\int d\vec r{\int d \vec t \hat{\psi} 
(\vec r,\vec t)\psi(\vec r,\vec t)}-\sum_{i=1}^N{\hat{\psi} 
(\vec r_i,\vec t_i)}}. 
\eeqa 
In this way we can rewrite eq. \ref{a2} as in Eq. \ref{partition_coins_2} 
with: 
%per i segni in \ref{a6} ricordare che f vale 0 o -1! 
\beqa\label{a6} 
{\mathcal H}\l(\psi,\hat{\psi}\r)=+\frac{1}{2}\int{d\vec r_1 d\vec r_2 
\psi(\vec r_1,\vec t_1)V_{2b}(\vec{r}_1-\vec{r}_2)\psi(\vec r_2,\vec t)} 
-i\int{d\vec r d\vec t \hat{\psi}(\vec r,\vec t)}\psi(\vec r,\vec t)\\ \nonumber 
-\log\l(\zeta(\hat{\psi})\r)-\frac{1}{2V}\int d\vec r_1{\int d\vec  
r_2{\psi(\vec r_1)\psi(\vec r_2) f^{(1)}(\vec r_1,\vec r_2)}} \\ \nonumber 
 -  \frac{1}{3V}\int d\vec r_1{\int d\vec r_2{\int d\vec r_3{\psi(\vec r_1) 
\psi(\vec r_2)\psi(\vec r_3)f^{(1)}(\vec r_1,\vec r_2)f^{(1)}(\vec r_2,\vec r_3)f^{(1)} 
(\vec r_1,\vec r_3)}}}\\ \nonumber 
- \frac{1}{2V}\int d\vec r_1{\int d\vec r_2{\psi(\vec r_1,\vec t_1) 
\psi(\vec r_2,\vec t_2)f^{(2)}(\vec r_1,\vec r_2)}} 
\eeqa 
where we have called $\zeta(\hat{\psi})$ the polymeric partition function, 
which reads: 
\beqa\label{a7} 
\zeta(\hat{\psi})\equiv\l(\int{\prod_i d\vec r_i \delta\l(|\vec{r}_{i+1} 
-\vec{r}_i|-1\r)e^{-i\sum_{i=1}^N{\hat{\psi}(\vec r_i,\vec t_i)}}}\r)^{\frac{1}{N}}. 
\eeqa 
The saddle point equations read $\frac{\delta {\mathcal H}}{\delta \psi} 
=\frac{\delta {\mathcal H}}{\delta\hat{\psi}}=0$. 
Solving this equation instead of finding the full partition function is
effectively a mean field approximation (the path we follow is
reminiscent of that used in Ref. \cite{garel}). 
In this form this equation is still not 
solvable explicitly. As in the text, we make the further approximation
$\psi(\vec r,\vec t)\equiv\rho\phi(\vec t)\equiv\psi(\vec t)$ (and consequently  
from the saddle point equations one finds 
$\hat{\psi}(\vec r,\vec t)\equiv\hat{\rho} 
\hat{\phi}(\vec t)\equiv\hat{\psi}(\vec t)$ ). One obtains
$\zeta(\hat{\psi})=\int{d\vec t \delta\l(|\vec{t}|-1\r)e^{-i{\hat{\psi}(\vec t)}}}$  
and the equation $\frac{\delta {\mathcal H}}{\delta\hat{\psi}}=0$ takes on the form of 
a self-consistent equation for $\psi$ and $\hat{\psi}$: 
\beqa\label{a8} 
\psi(\vec r,\vec t)=\rho\frac{e^{-i{\hat{\psi}(\vec t)}}} 
{\int {d\vec t \delta\l(|\vec{t}|-1\r)e^{-i{\hat{\psi}(\vec t)}}} } 
\eeqa 
or equivalently: 
\beqa\label{a9} 
-i\hat{\psi}(\vec t)=\log{\l(\frac{\langle e^{-i{\hat{\psi}(\vec t)}} 
\rangle\psi(\vec r,\vec t)}{\rho}\r)} 
\eeqa 
where with $\langle\cdot\rangle$ we denote ensemble averaging with respect to the 
measure $\int{d\vec t \delta\l(|\vec{t}|-1\r)\cdot}$. To we recover the  
free energy in Eqs. \ref{mean_field_functional_coins}, we need to recall also 
the formula for the mean excluded volume of two coins, with axis $\vec t_1$  
and $\vec t_2$, and radius $R_0$ which is due to Onsager\cite{onsager} and reads: 
\beq\label{a10} 
\frac{1}{2V}\int d\vec r_1{\int d\vec r_2  
f^{(2)}(\vec r_1,\vec r_2;\vec t_1,\vec t_2)}= 
-2\pi R_0^3\l(1-\l(\vec t_1\cdot\vec t_2\r)^2\r)^{\frac{1}{2}}. 
\eeq 
Inserting Eqs. \ref{a7} into Eq. \ref{partition_coins_2}, and using Eq. \ref{a9}, we 
obtain Eq. \ref{mean_field_functional_coins}. 
%\end{document}

\newpage

%\begin{figure}
%\centerline{\psfig{figure=defbravais.eps,angle=270,width=2.5in}}
%\caption{Plot of the angle defining a two dimensional Bravais lattice.}
%\label{definition_bravais_2d3d}
%\end{figure}

\newpage 

\begin{figure}
\centerline{\psfig{figure=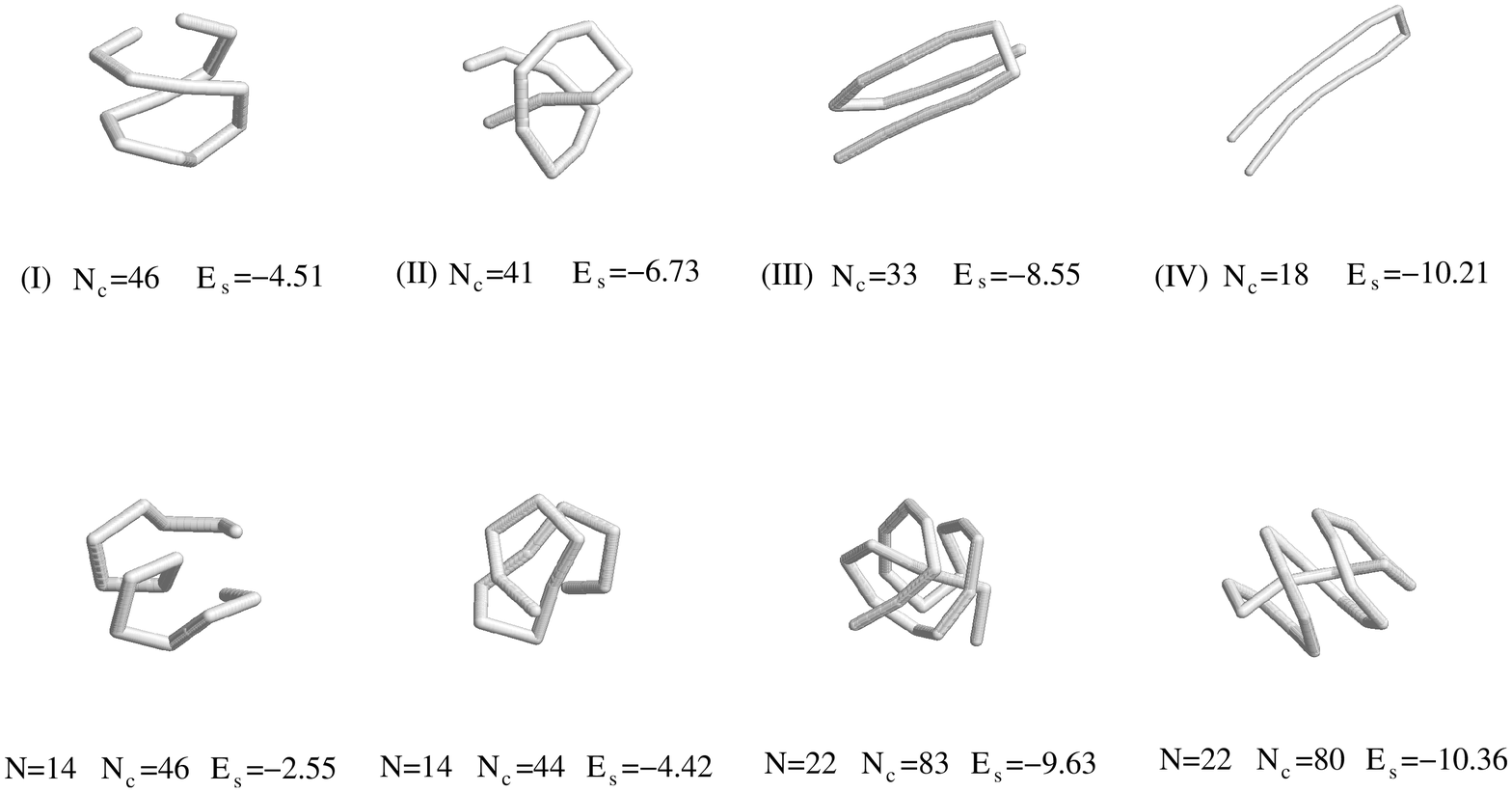,width=6.5in}}
\label{stiff_configurations}
\caption{In this figure we show the ground states obtained for a semiflexible
chain of increasing stiffness. The notes under each configurations indicate
its number of contacts and the value of the stiffness term $E_s$
(see the text for its definition).
The labels I, II, III and IV are discussed in the text. Conformations
in the first row have chain length $N=14$.}
\end{figure}

\begin{figure}
\centerline{\psfig{figure=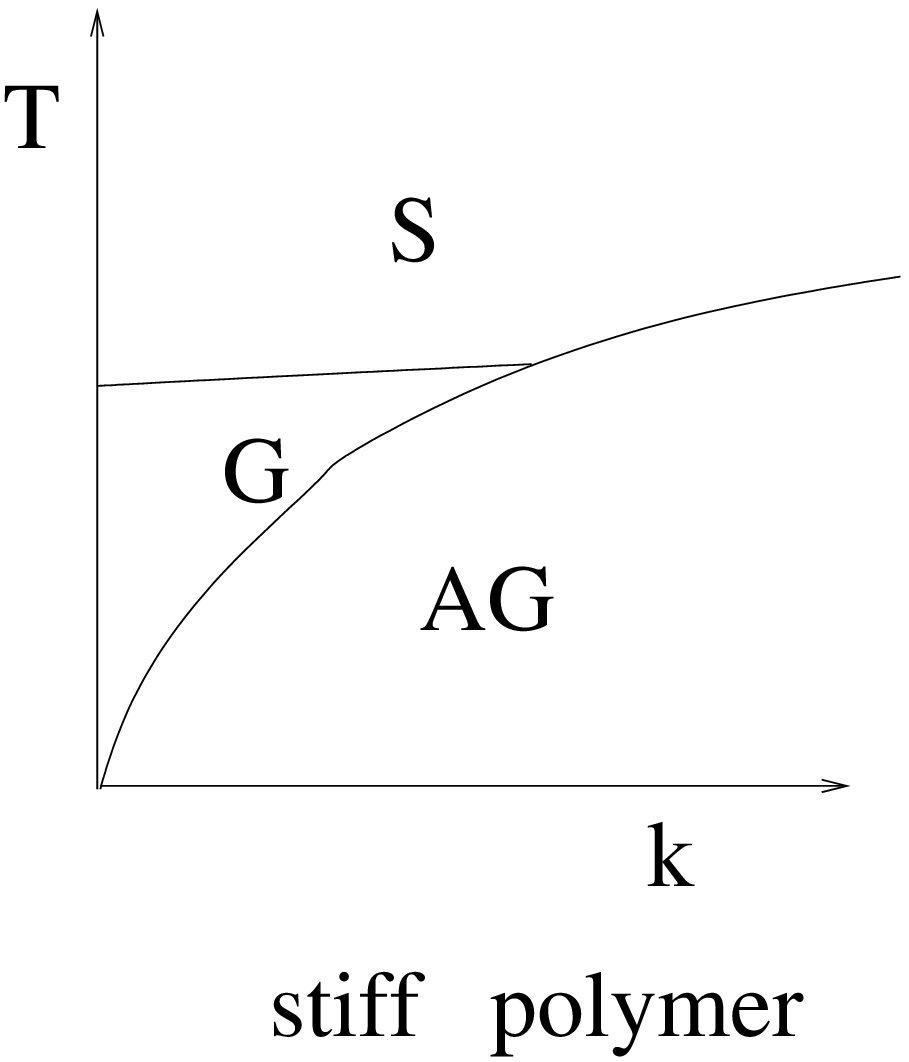,width=3.in}}
\caption{ Schematic phase diagram of a stiff polymer. The phases  are
S=Swollen, G=globule, and AG= Asymmetric globule.}
\label{stiff}
\end{figure}

\begin{figure}
\psfig{figure=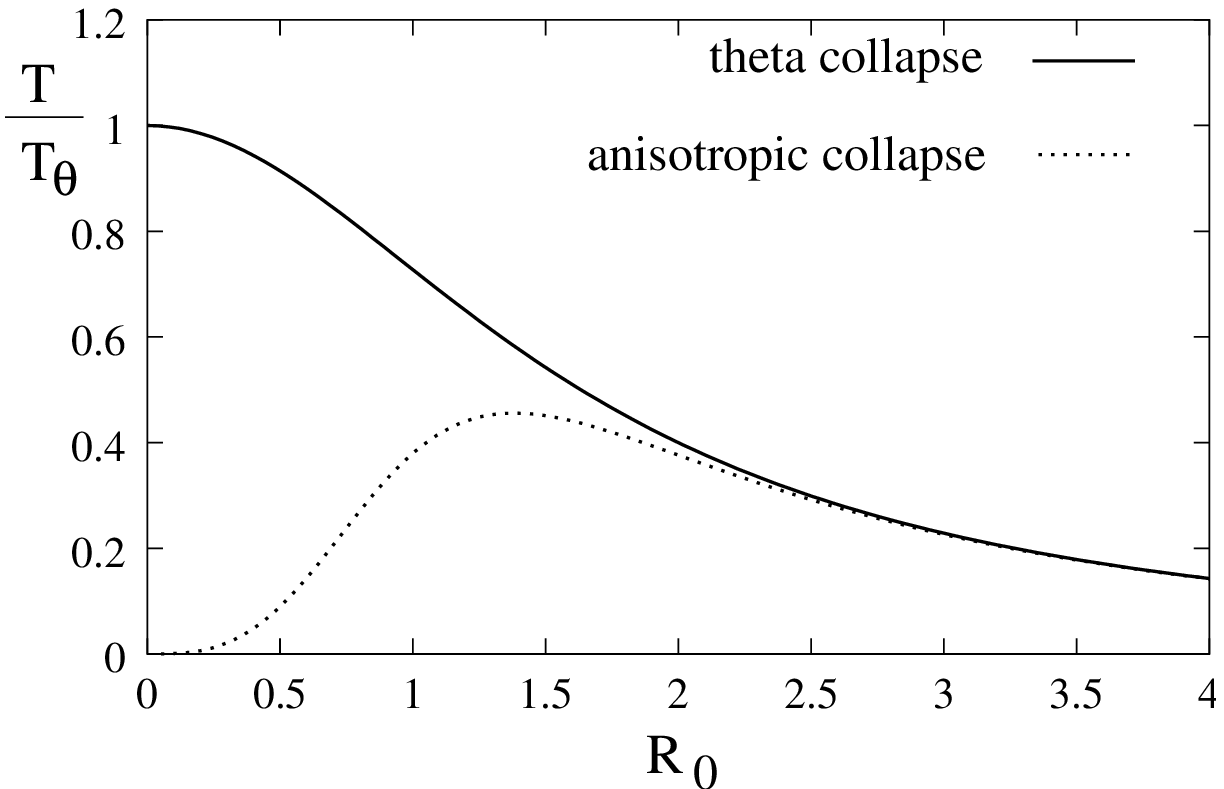,width=3.5in}
\psfig{figure=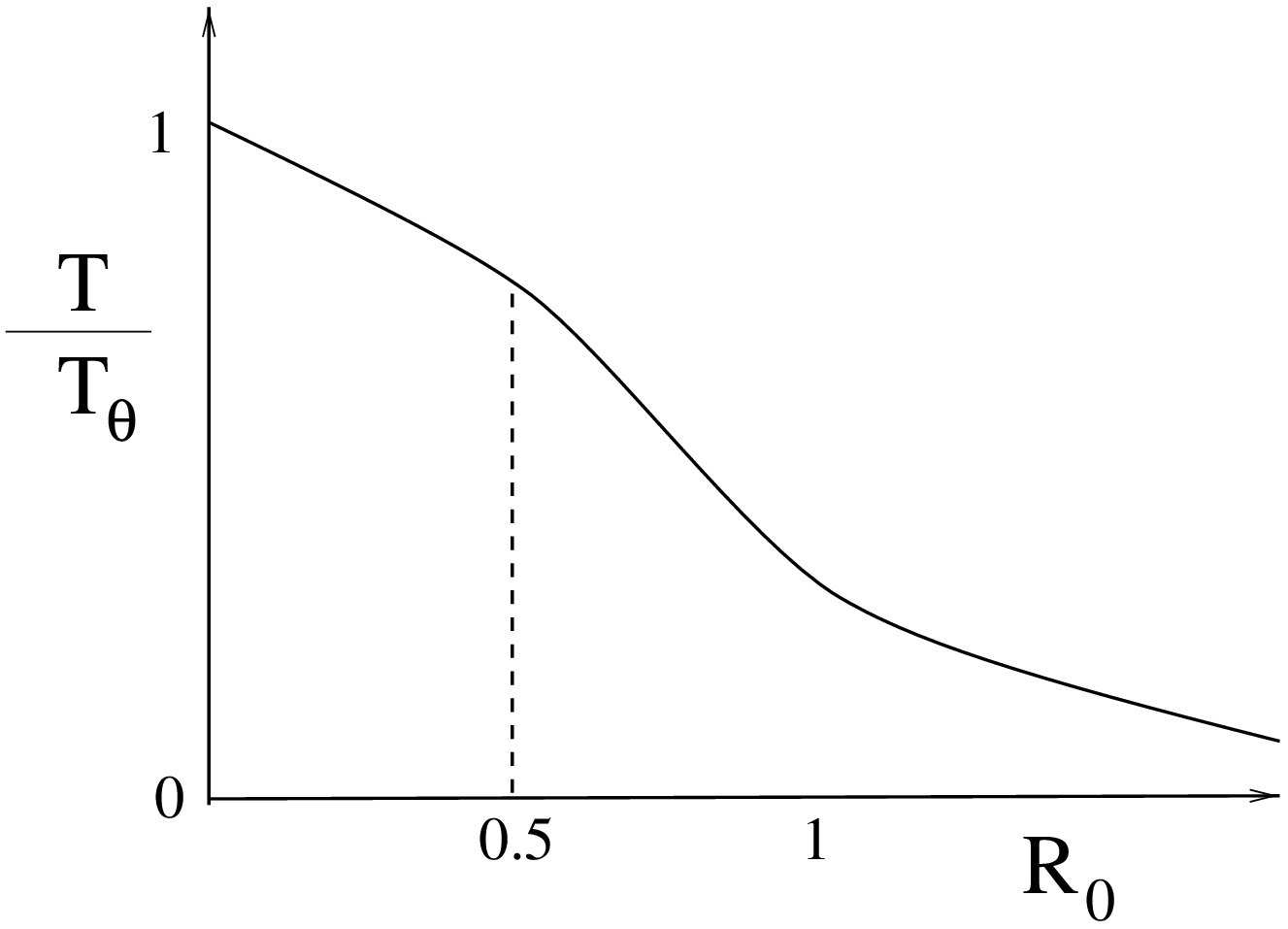,angle=270,width=3.5in}
\caption{Schematic phase diagrams  of a chain of coins 
(top panel) and of a tube with a three body constraint 
(bottom panel) in the $(T,R_0)$ plane obtained in the
mean field approximation.}
\label{phase_diagram_mean_field}
\end{figure}

\begin{figure}
\psfig{figure=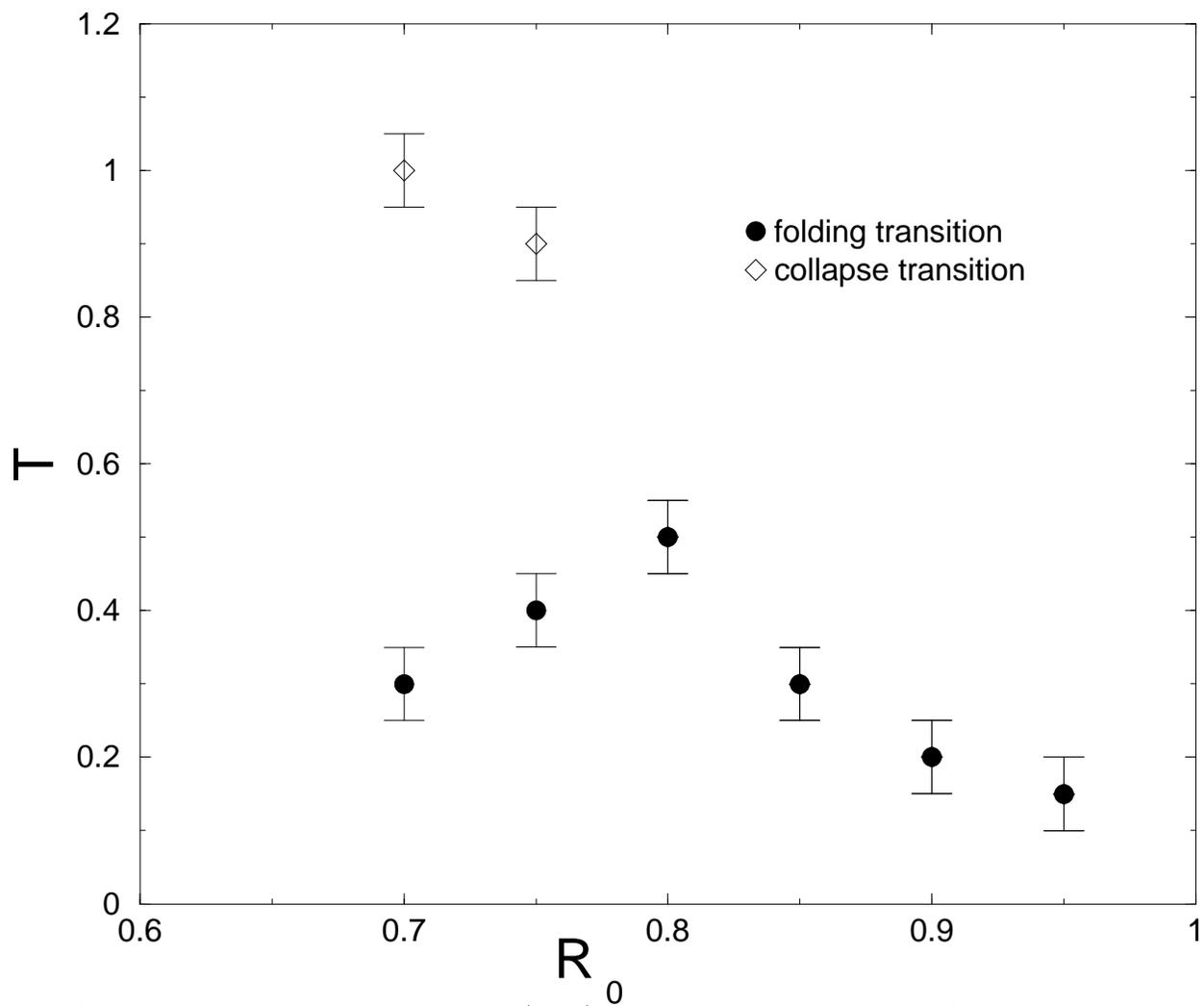,angle=270,width=6.5in}
\caption{Phase diagram for a thick polymer in the $(T,R_0)$ plane obtained 
with Monte-Carlo simulations. The points represent peaks in the specific heat curve for 
$N=41$.}
\label{phase_diagram_mc}
\end{figure}

%two figures with specific heat--gyration radius
\begin{figure}
\centerline{\psfig{figure=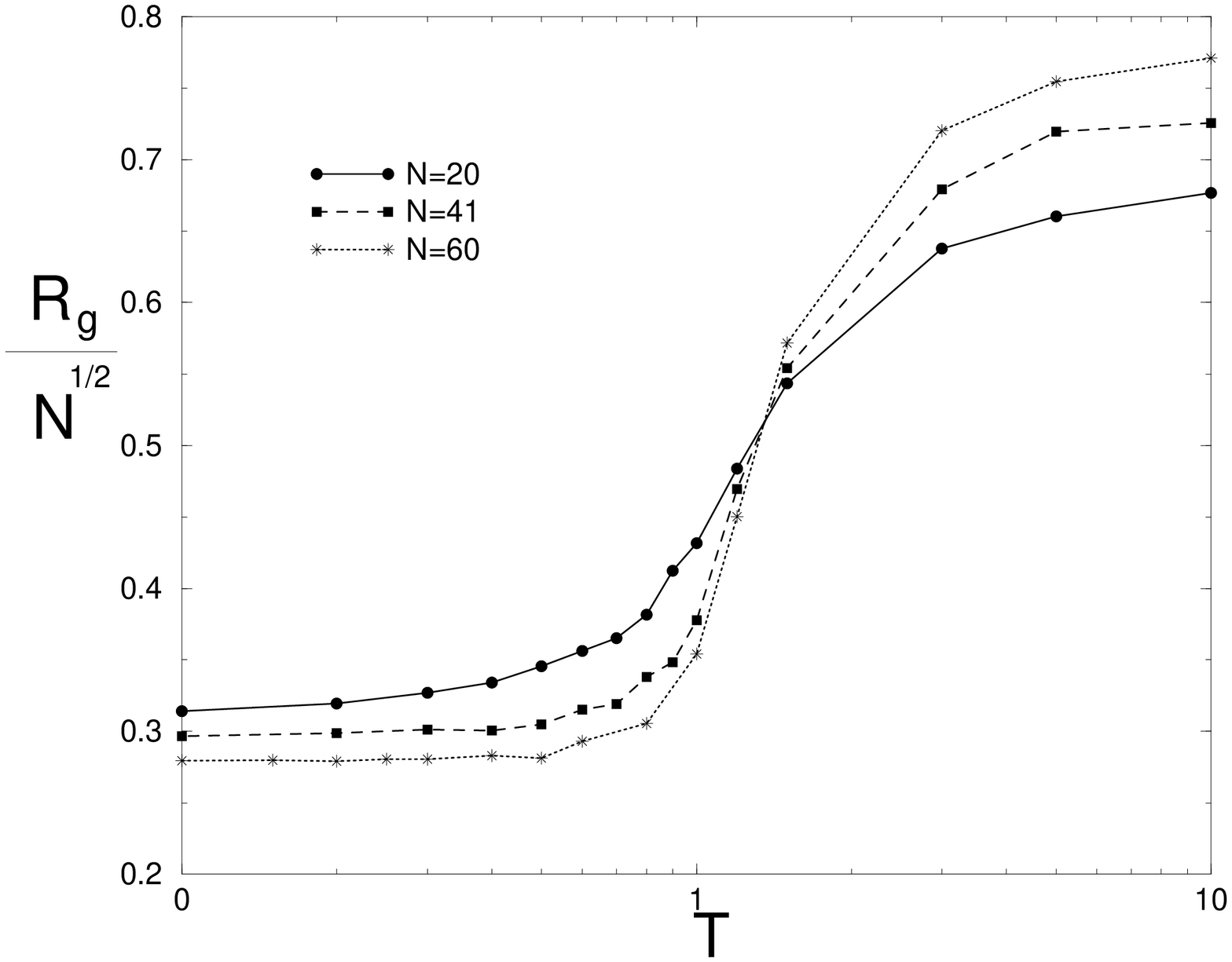,angle=270,width=6.in}}
\centerline{\psfig{figure=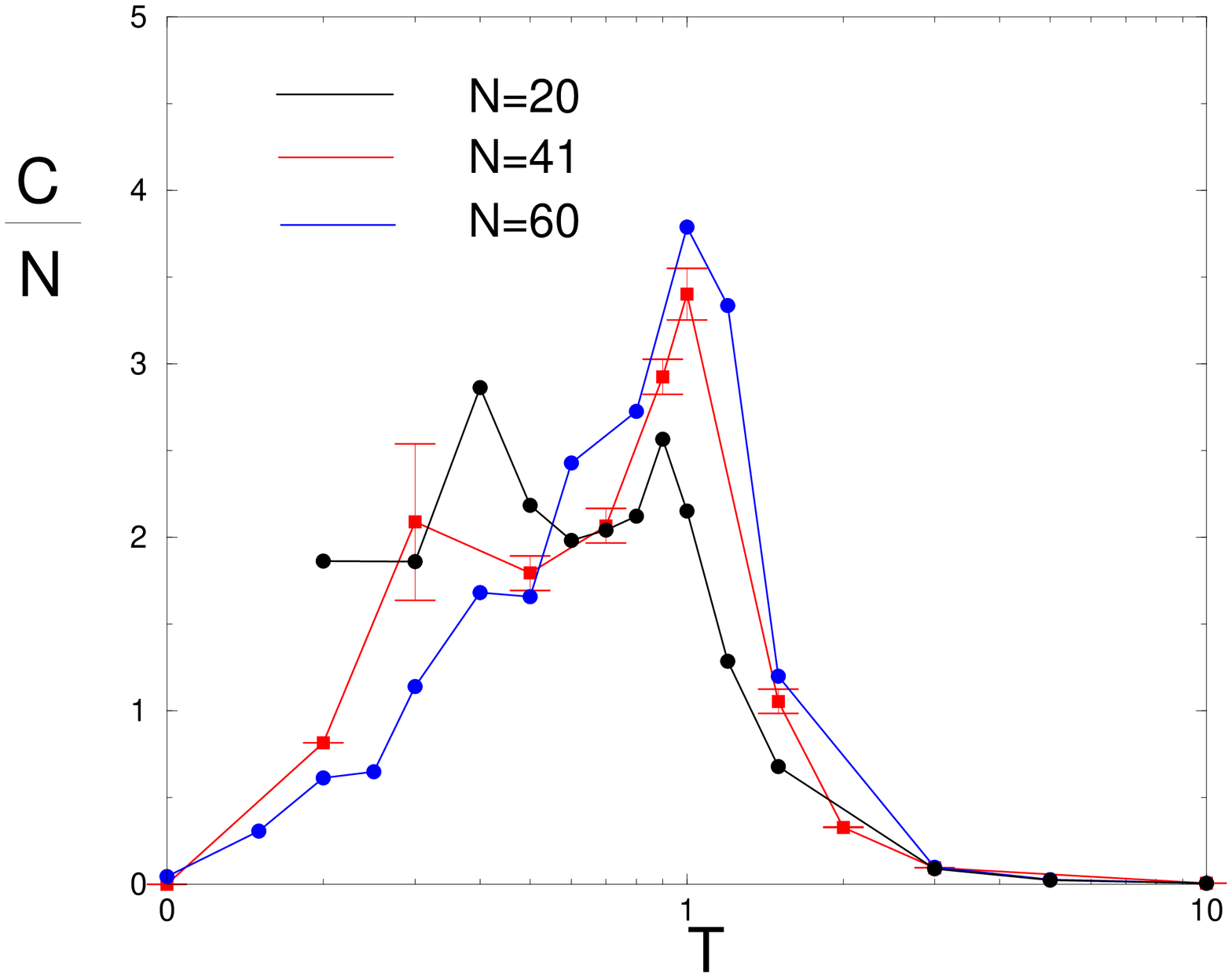,angle=270,width=6.in}}
\caption{Plots of radius of gyration (top) and specific heat (bottom) 
versus temperature for a thick polymer
with $R_0=0.7$. The error bars in the radius of gyration estimate are comparable to
the point size. For clarity, we have shown error bars in the specific heat
only for $N=41$, for which we have carried out the most extensive
simulations.}
\label{gir_07}
\end{figure}

\begin{figure}
\centerline{\psfig{figure=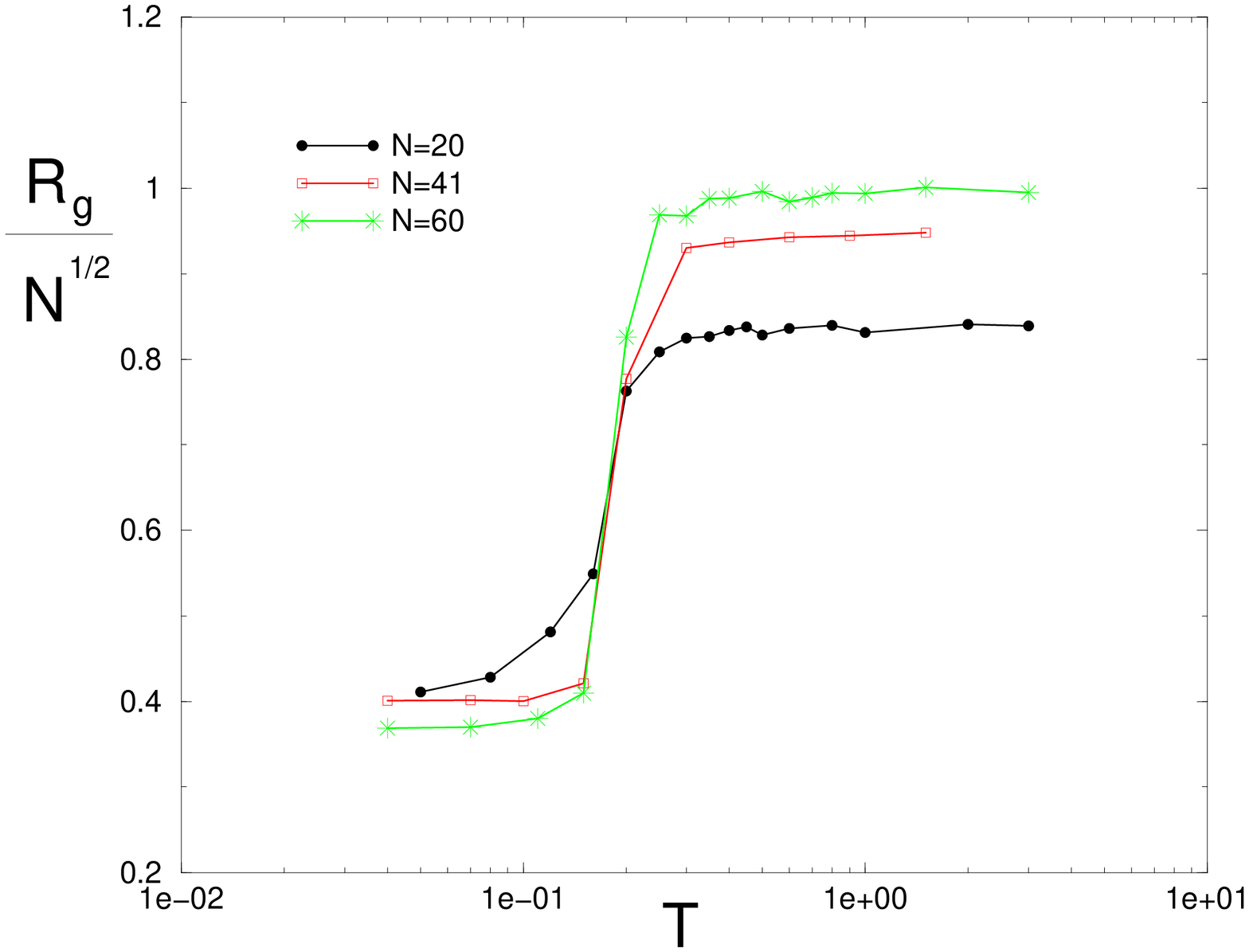,angle=270,width=6.in}}
\centerline{\psfig{figure=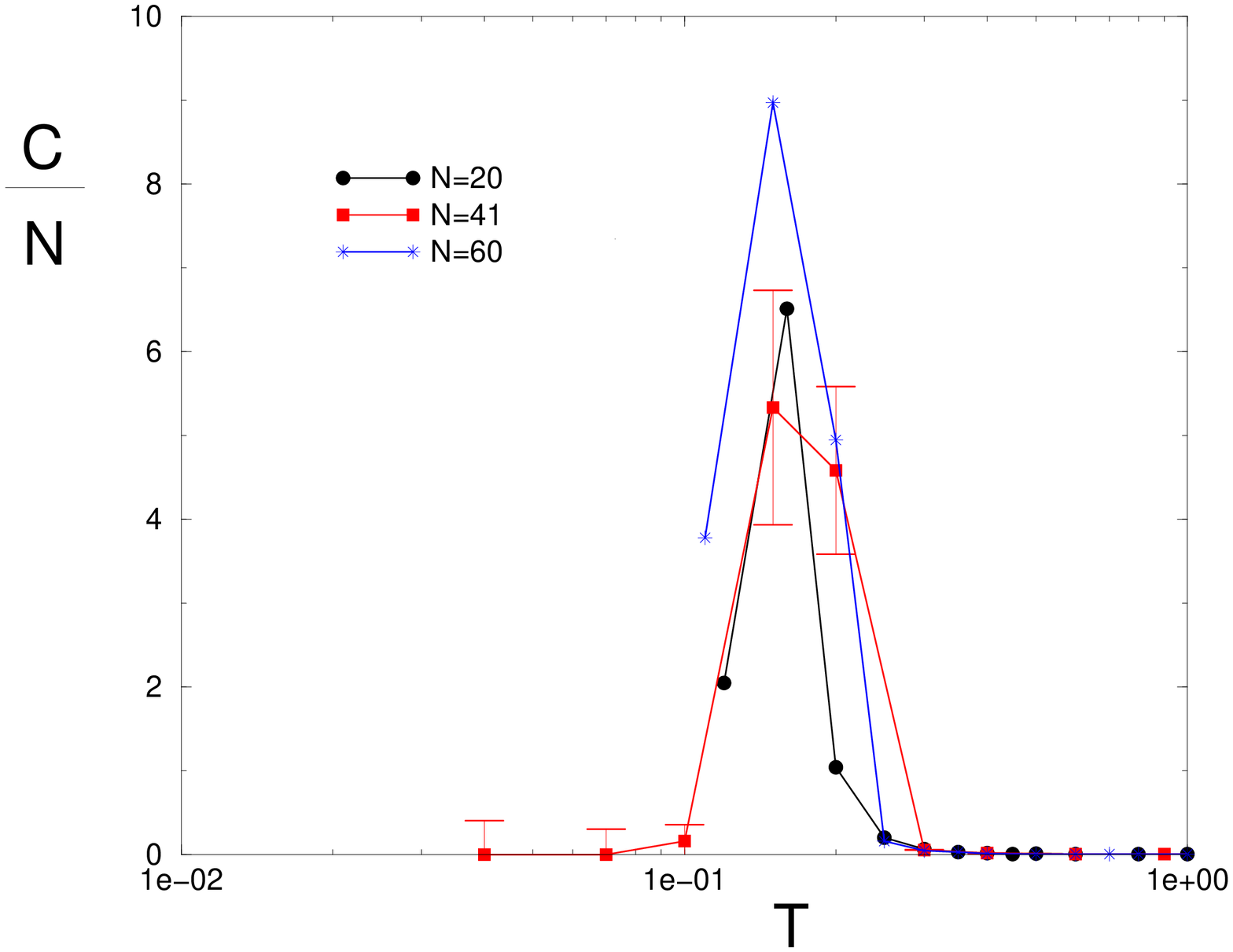,angle=270,width=6.in}}
\caption{Plots of radius of gyration (top) and specific heat (bottom) 
versus temperature for a thick polymer
with $R_0=0.95$. The error estimates are as described in the caption 
of Fig. \ref{gir_07}}
\label{gir_095}
\end{figure}

\begin{figure}
\psfig{figure=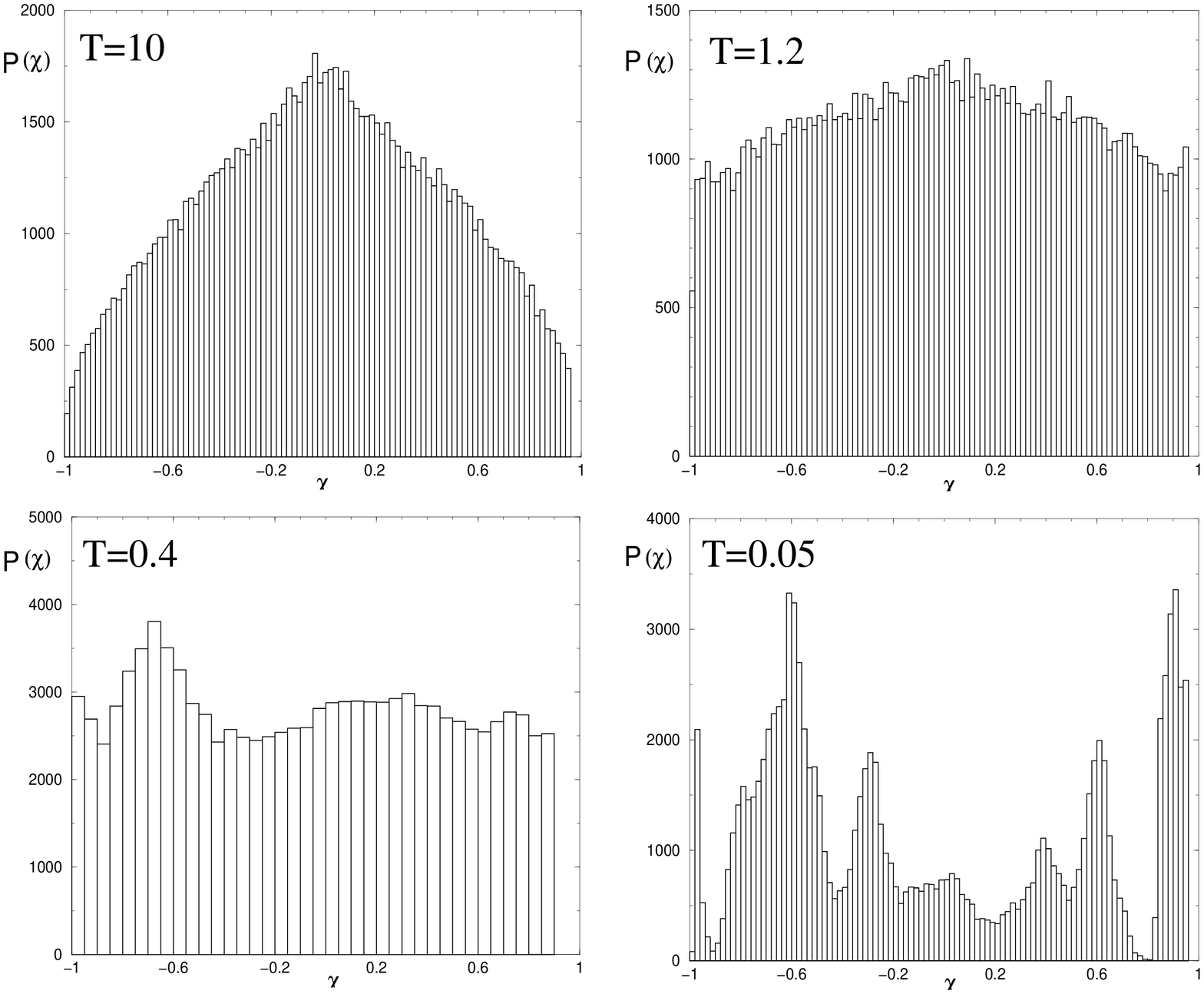,width=6.5in}
\caption{Chirality probability distribution (unnormalized) 
from Monte-Carlo simulations
for a thick polymer with $R_0=0.7$.}
\label{chi_07}
\end{figure}

\begin{figure}
\psfig{figure=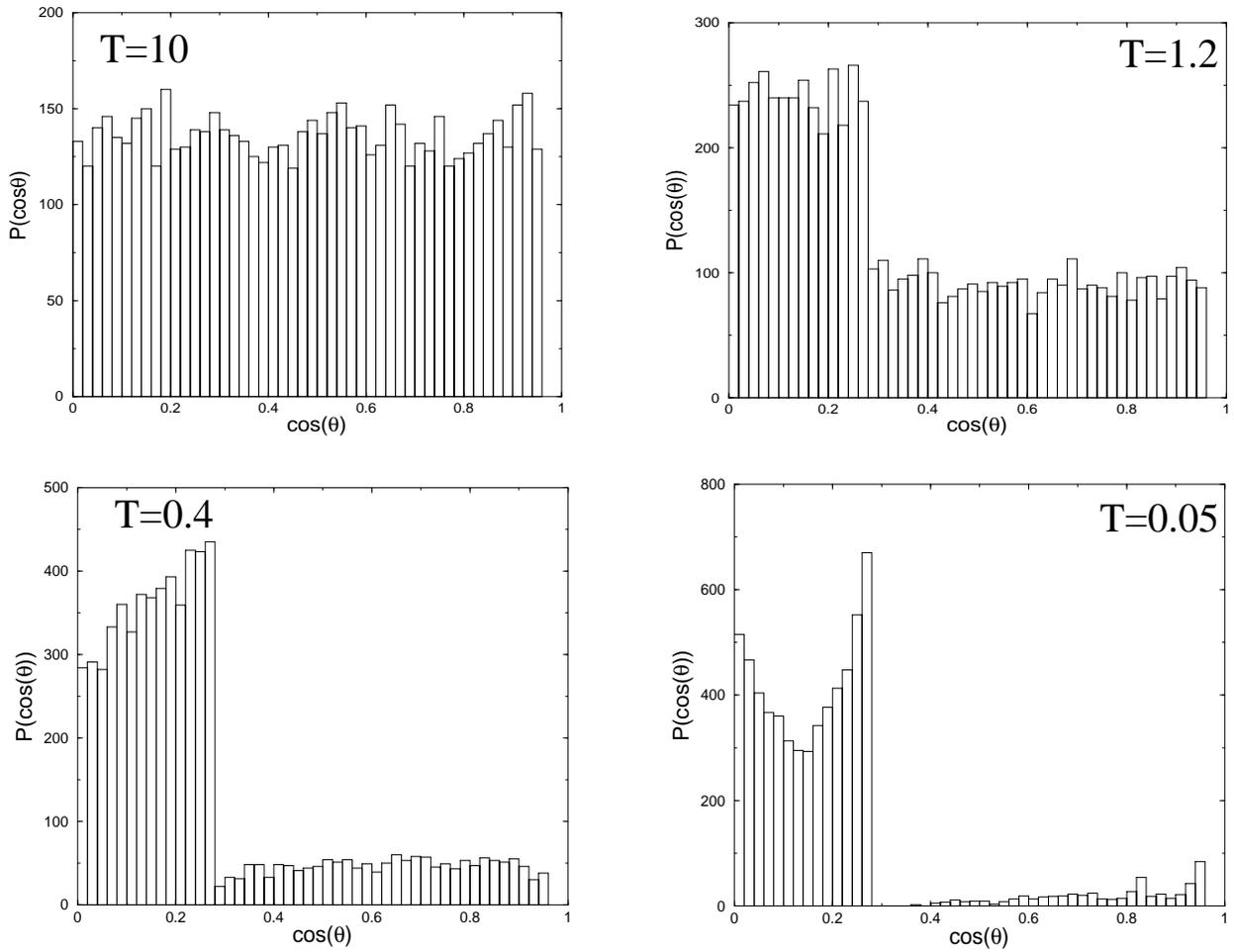,width=6.5in}
\caption{Unnormalized 
probability distribution of $\cos{(\theta)}$ for $R_0=0.7$.
The $x$-axis scale begins at $0$ because $R_0=0.7$ yields
a constraint on the local thickness that $\cos{\theta}$ 
must be greater than $-0.02\ldots$ and virtually no points have negative
$\cos{\theta}$.}
\label{theta_07}
\end{figure}

\begin{figure}
\psfig{figure=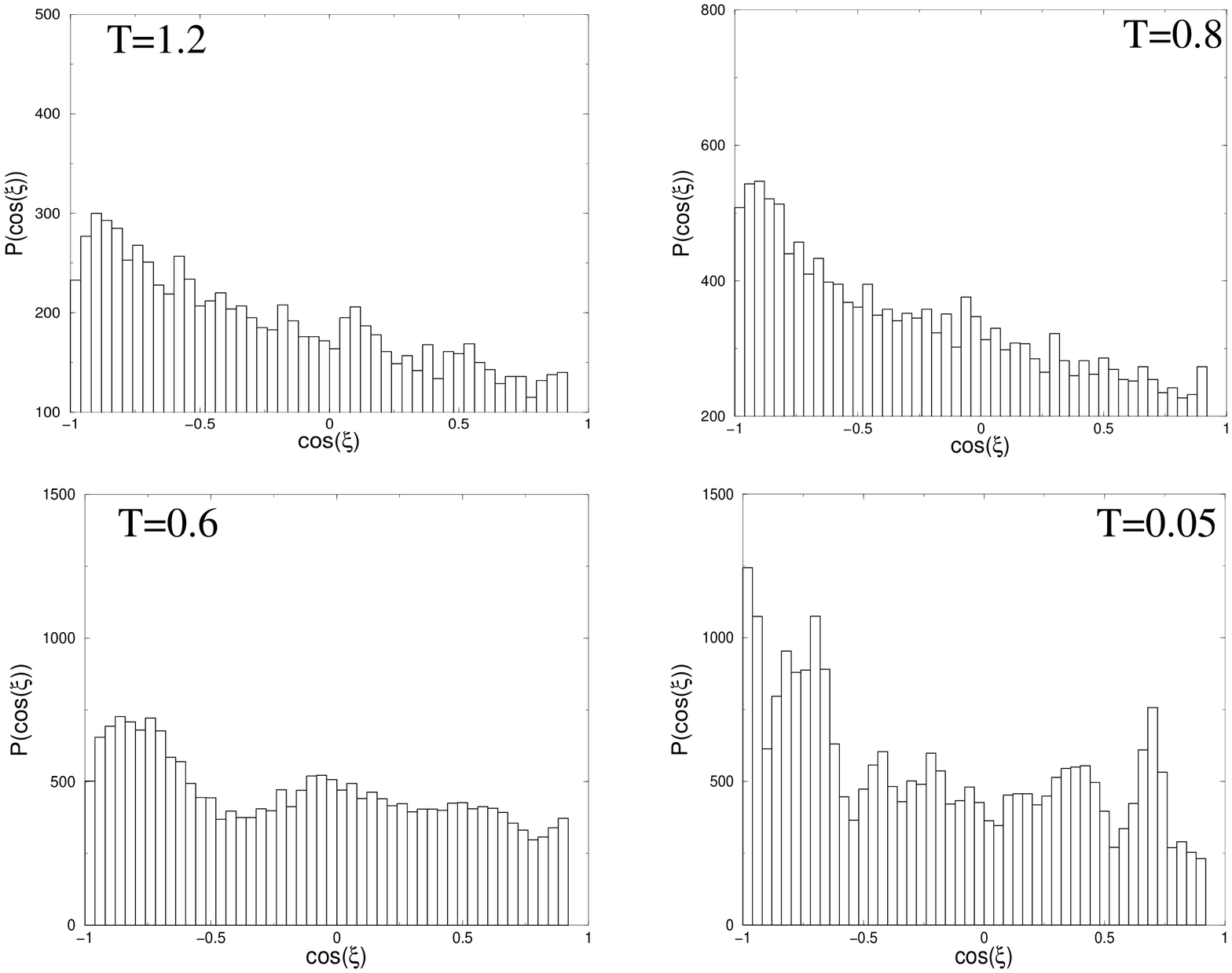,width=6.5in}
\caption{Unnormalized probability distribution of $\cos{(\xi)}$ for $R_0=0.7$.}
\label{xi_07}
\end{figure}

\begin{figure}
\psfig{figure=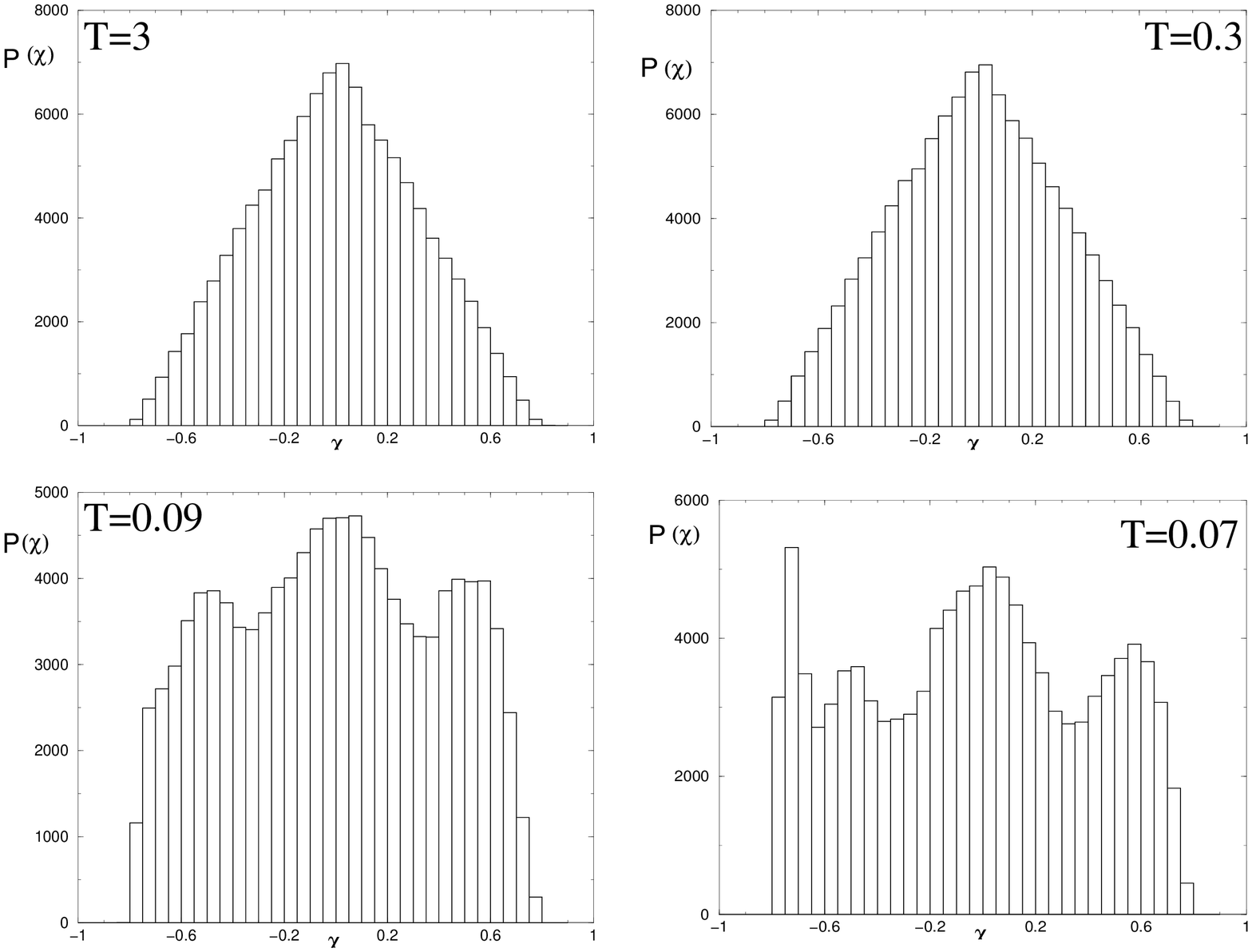,width=6.5in}
\caption{Chirality probability distribution (unnormalized) 
from Monte-Carlo simulations
for a thick polymer with $R_0=0.95$.}
\label{chi_095}
\end{figure}

\begin{figure}
\psfig{figure=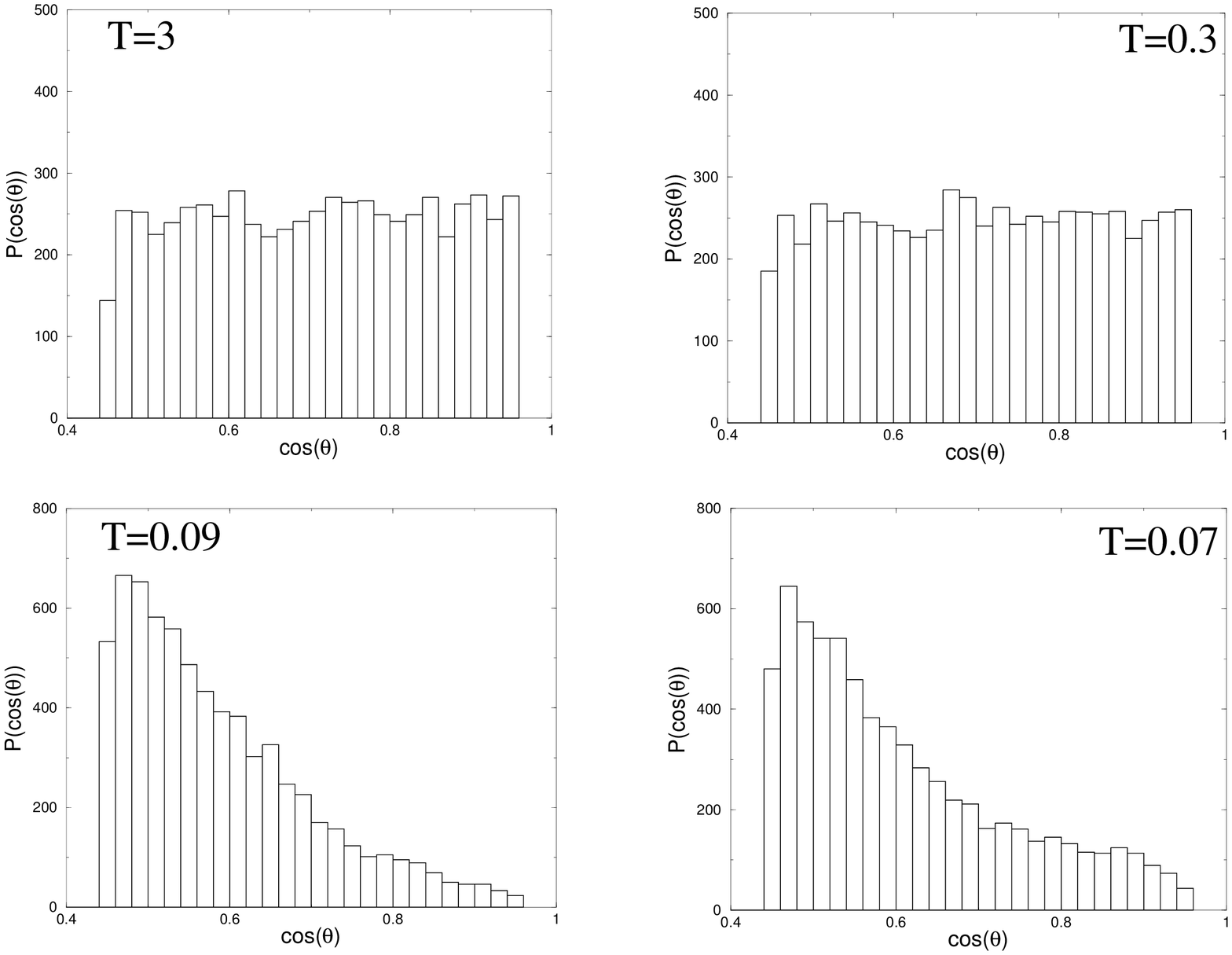,width=6.5in}
\caption{Unnormalized probability 
distribution of $\cos{(\theta)}$ for $R_0=0.95$.}
\label{theta_095}
\end{figure}

\begin{figure}
\psfig{figure=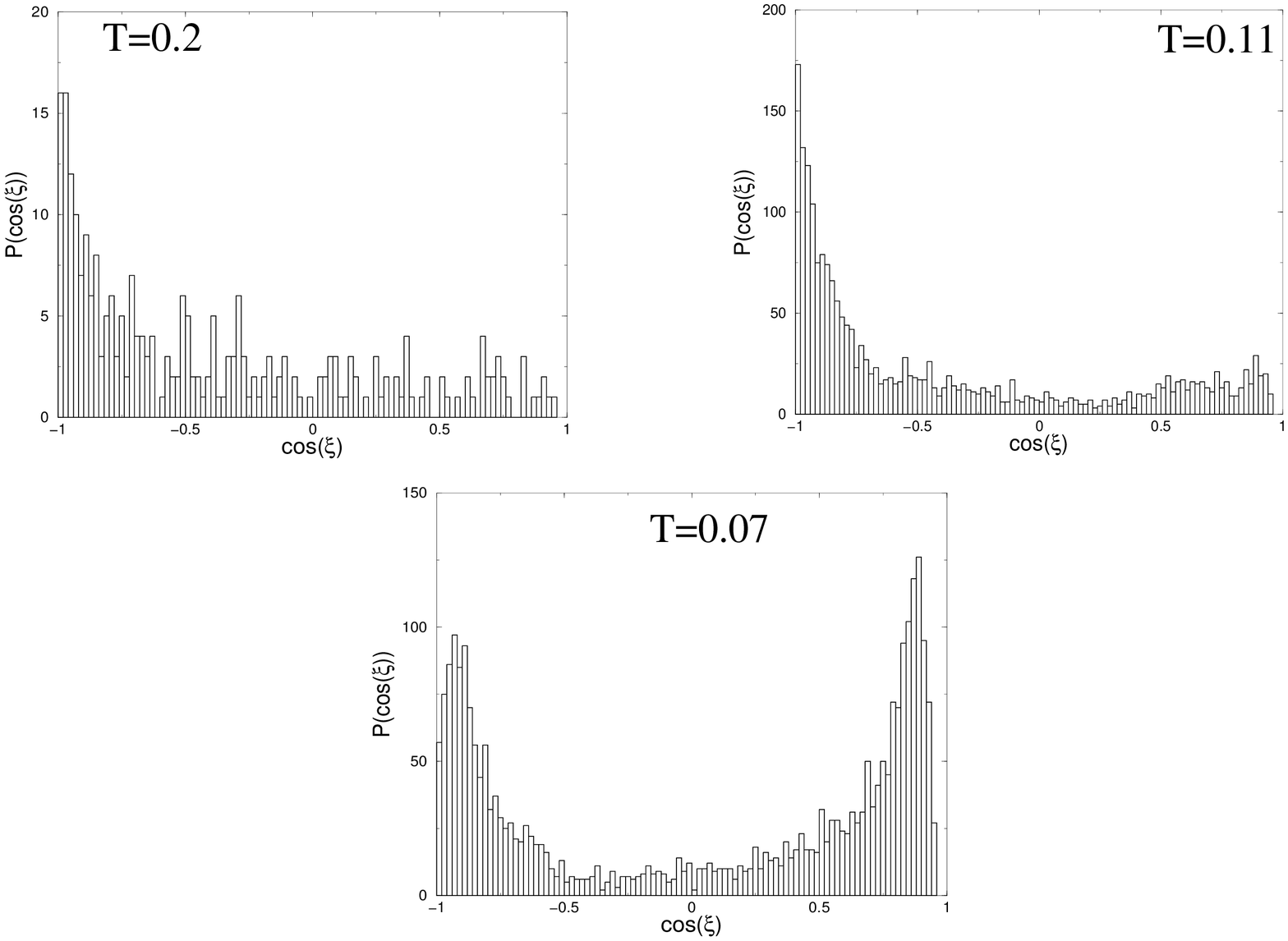,width=6.5in}
\caption{Unnormalized 
probability distribution of $\cos{(\xi)}$ for $R_0=0.95$.}
\label{xi_095}
\end{figure}


\begin{thebibliography}{99}
\bibitem{note11} In a different context, recent studies\cite{frenkel,benedek,frenkel1} 
have underscored the importance of the range of the attractive interactions 
in determining the stability of the fluid-fluid coexistence curve in a 
system of hard spheres and the importance of anisotropy in determining the 
phase diagram of protein solutions.
\bibitem{note12} Note that models different from ours 
but incorporating anisotropy of the chain backbone
have been proposed for discrete polymer chains in the past 
(see Refs. \cite{odijk,khokhlov,odijk1,odijk2,khokhlov1} for some
examples). Our  model  however allows one to take the
continuum limit without introducing further cut-offs in the theory.
\bibitem{PHI4} One can formally map the model in eq. (\ref{ham}) into a
field theory of an $n$-component field, $\vec{\phi}(\vec{x})$. In the
grand canonical formulation with a fugacity per unit length, $r$,
the Hamiltonian of the field theory is given by  
\begin{equation}
H_n(\{\phi\})= \int d^dx
\{\frac{1}{2}\partial_{\mu}\vec{\phi}\partial_{\mu}\vec{\phi} + 
r\vec{\phi}^2 + \sum_{m}\frac{v_m}{m!}(\vec{\phi}^2)^m\} 
\end{equation}
Correlation functions for the corresponding polymer model are
calculated from the $\vec{\phi}$ correlation in the $ n \rightarrow 0$ 
limit \cite{PhysLettA,ZinnJustin}. 
The renormalizability of such a field theory is well known
and it is guaranteed that, from a perturbative expansion, 
one can extract results
which depend only on a finite number of parameters.
\bibitem{note13} 
An interesting problem for thick polymers in the continuum 
limit is the  measure associated with the curve corresponding
to the viable centerline
of a thick tube. Indeed, for the Edwards model, the first term
in the effective energy in Eq. (1) is used to define a measure on the
ensemble of all continuous curves embedded in three-dimensional
space, the Wiener measure.  It is well known that this gives a
non-zero weight to all curves which are continuous but not
differentiable ($C^0$ curves in the standard notation)). In
particular, the curves for which the tangent can be defined are a
subset of zero measure in this ensemble. It can be proved rigorously
(see e.g. Ref. \cite{gonz}) that if a curve  is a viable centerline
for a tube of non zero thickness, then it must satisfy more
stringent smoothness requirements, namely it should be $C^{1,1}$,
i.e. it is differentiable and its derivative is  a Lipschitz function
(see \cite{gonz} and Refs. therein).  An interesting problem, which we 
do not consider here,
is the definition of a suitable measure in the continuum limit for such
thick polymer centerlines.  One possibility is, of course, to add, to the
first term in Eq. (1), the three body potential introduced in this
work defined for all possible triplet radii, but this, for the
reason just mentioned, does not lend itself to a straightforward
perturbative formulation.
\bibitem{hull} We call a tube  space
filling if the smallest convex set which contains it (its convex hull) 
is simply connected, i.e. has no holes at its interior.
\bibitem{note1}{Looking at the conformation with the chain
not drawn it is evident that the balls are sitting on the sites of
a regular lattice, possibly {\em fcc}.}
\bibitem{note2}{ The derivation of Eq. \ref{a4} implies
neglecting the chain constraint in the interaction part.
It is possible but more cumbersome to take into account the chain
constraint properly also here\cite{domb}. For our purposes, however 
it suffices to show that one obtains a term of second order in $\rho$ in the
free energy. }
\end{thebibliography}
\end{document}